\def\Im{{\text{Im}}\,}
\def\Re{{\text{Re}}\,}
\def\kF{k_{\text{F}}}
\def\vF{v_{\text{F}}}
\def\NF{N_{\text{F}}}
\def\me{m_{\text{e}}}

\def\sgn{{\text{sgn\,}}}
\def\be{\begin{equation}}
\def\ee{\end{equation}}
\def\bea{\begin{eqnarray}}
\def\eea{\end{eqnarray}}
\def\bse{\begin{subequations}}
\def\ese{\end{subequations}}

\documentclass[prb,twocolumn,showpacs,amsmath,amssymb,eqsecnum,preprintnumbers]{revtex4}
\usepackage{graphicx}
\usepackage{dcolumn}
\usepackage{bm}
\begin{document}
\preprint{KITP Preprint NSF-KITP-08-83}
\title{Theory of Helimagnons in Itinerant Quantum Systems IV: \\
       Transport in the Weak-Disorder Regime}
\author{T. R. Kirkpatrick$^1$, D. Belitz$^{2,3}$ and Ronojoy Saha$^2$}
\affiliation{$^{1}$ Institute for Physical Science and Technology,
                    and Department of Physics, University of Maryland, College Park,
                    MD 20742, USA\\
             $^{2}$ Department of Physics and Institute of Theoretical Science,
                    University of Oregon, Eugene, OR 97403, USA\\
             $^{3}$ Kavli Institute for Theoretical Physics, University of
                    California, Santa Barbara, CA 93106, USA}
\date{\today}

\begin{abstract}
We apply a recent quasiparticle model for the electronic properties of metallic
helimagnets to calculate the transport properties of three-dimensional systems
in the helically ordered phase. We focus on the ballistic regime $\tau^2 T
\epsilon_{\text{F}} >> 1$ at weak disorder (large elastic mean-free time
$\tau$) or intermediate temperature. In this regime, we find a leading
temperature dependence of the electrical conductivity proportional to $T$. This
is much stronger than either the Fermi-liquid contribution ($T^2$) or the
contribution from helimagnon scattering in the clean limit ($T^{5/2}$). It is
reminiscent of the behavior of non-magnetic {\it two-dimensional} metals, but
the sign of the effect is opposite to that in the non-magnetic case.
Experimental consequences of this result are discussed.
\end{abstract}


\maketitle

\section{Introduction}
\label{sec:I}

The electrical transport properties of metals have given rise to various
surprises over the last thirty years. Within a nearly-free electron model with
quenched or static disorder, and to lowest order in the impurity concentration,
the Boltzmann equation is exact and yields the familiar Drude expression for
the electrical conductivity,
\be
\sigma_0 = ne^2\tau/m_e\ ,
\label{eq:1.1}
\ee
with $n$ the electron number density, $e$ the electron charge, $m_{\text{e}}$
the effective electron mass, and $\tau$ the elastic mean-free time between
collisions, which is weakly temperature dependent. Corrections to this result,
in an expansion in the small parameter $1/\epsilon_{\text{F}}\tau$, with
$\epsilon_{\text{F}}$ the Fermi energy, turned out to be very interesting. To
analyze them, one needs to distinguish, in the thermodynamic limit, between the
diffusive regime at strong disorder or low temperature, $T\tau \ll 1$, and the
ballistic regime at weak disorder or intermediate temperature, $T\tau \gg 1$.
(The latter regime should not be confused with ballistic transport in
mesoscopic systems, where the mean-free path is large compared to the system
size.) In three-dimensional (3D) simple metals in the diffusive regime, the
leading correction is nonanalytic in the temperature $T$,
\cite{Gorkov_Larkin_Khmelnitskii_1979, notation_footnote}
\bse
\label{eqs:1.2}
\be
\delta\sigma_{\text{WL}} \propto
\sigma_0\,\frac{(T\tau)^{1/2}}{(\epsilon_{\text{F}}\tau)^2}\quad \text{(3D,
diffusive)}\quad.
\label{eq:1.2a}
\ee
The sign of this effect is positive, which reflects a negative $T$-independent
contribution that is non-universal (i.e., depends on an ultraviolet cutoff).
The effect thus is localizing, i.e., it decreases the conductivity compared to
the Drude value. In two-dimensional (2D) systems the effect is even more
dramatic, \cite{Abrahams_et_al_1979}
\be
\delta\sigma_{\text{WL}} \propto
\sigma_0\,\frac{\ln(T\tau)}{\epsilon_{\text{F}}\tau}\quad \text{(2D,
diffusive)}\quad.
\label{eq:1.2b}
\ee
\ese
These results have been reviewed in Ref.\ \onlinecite{Lee_Ramakrishnan_1985}.
The logarithmic divergency in 2D perturbation theory signals a breakdown of
transport theory, and the behavior at $T=0$ is insulating, albeit very weakly
so. These effects in non-interacting electron systems in general, and the
logarithmic temperature dependence in 2D in particular, are known as ``weak
localization''. They can be understood in terms of constructive interference in
the electron-impurity scattering process,\cite{Bergmann_1984} or in terms of
the exchange of certain soft or massless diffusive modes (either ``Cooperons'',
or ``diffusons'') between electrons.

Taking into account the screened Coulomb interaction between electrons leads to
additional effects, and considerably enhances the complexity of the
calculations. In the absence of quenched disorder, Fermi-liquid theory
accurately describes the behavior and leads to a conductivity given by the
Drude formula
\bse
\label{eqs:1.3}
\be
\sigma_C = ne^2 \tau_C/m_{\text{e}},
\label{eq:1.3a}
\ee
with a Coulomb scattering rate proportional to $T^2$,\cite{Schmid_1974}
\be
1/\tau_C = \frac{\pi^3}{8}\,\frac{T^2}{\epsilon_{\text{F}}}\ .
\label{eq:1.3b}
\ee
\ese
In the presence of both disorder and a Coulomb interaction, and in the
diffusive regime, the leading corrections in the diffusive regime are
qualitatively the same as those shown in Eqs.\
(\ref{eqs:1.2}).\cite{Altshuler_Aronov_Lee_1980, Altshuler_Aronov_1984}
However, the physics behind the effects is different, and this is reflected in,
e.g., the sensitivity of the results to an external magnetic field. These
effects are often referred to as ``Altshuler-Aronov effects''. In the ballistic
regime, $T\tau \gg 1$, the same problem has been analyzed for 2D systems by
Zala et al. \cite{Zala_Narozhny_Aleiner_2001} They found
\bse
\label{eqs:1.4}
\be
\delta\sigma_{\text{AA}} \propto \sigma_0\,T/\epsilon_{\text{F}}\quad
\text{(2D, ballistic)}\quad.
\label{eq:1.4a}
\ee
The sign of the correction depends on the interaction strength, but generically
it is localizing, as it is in the diffusive regime. More generally, the
conductivity correction can be written
\be
\delta\sigma_{\text{AA}}/\sigma_0 =
\frac{1}{\epsilon_{\text{F}}\tau}\,f(T\tau),
\label{eq:1.4b}
\ee
\ese
with $f(x\rightarrow 0) \propto \ln x$, and $f(x\rightarrow\infty) \propto x$
in 2D. The crossover between the two limits was also determined in Ref.\
\onlinecite{Zala_Narozhny_Aleiner_2001}, and a physical interpretation in terms
of scattering by Friedel oscillations was given.

The most appropriate interpretation of $\delta\sigma_{\text{AA}}$ is as the
result of a correction to the clean Fermi-liquid relaxation rate $1/\tau_C$. In
the ballistic regime, this correction is small compared to $1/\tau_C$ by a
factor of $1/T\tau \ll 1$. Adding this correction to $1/\tau_C$ and $1/\tau$
according to Matthiesen's rule leads, in the regime where $1/\tau_C \ll
1/\tau$, to Eq.\ (\ref{eq:1.4a}). Another possible interpretation of
$\delta\sigma_{\text{AA}}$ is as an interaction-induced temperature dependent
correction to $\sigma_0$. In the Coulomb case, $\delta\sigma_{\text{AA}}$ is
small compared to both $\sigma_C$ and $\sigma_0$, but we will see later that
this is not necessarily true if the correction is mediated by a different
interaction.

A general explanation of the physics behind these nonanalytic temperature
dependencies is as follows. If there are massless excitations that couple to
the relevant electronic degrees of freedom, then the lack of a mass in the
excitation propagator will lead to integrals that are singular in the infrared,
and these singularities are protected by a nonzero temperature $T>0$ (or
frequency in the zero-temperature limit), which leads to a nonanalytic
dependence on $T$. Notice that the massless excitation does not have to be of
non-electronic origin: with a proper classification of modes they can
themselves be electronic in nature without leading to double counting. For
instance, in the case of the weak-localization singularities the relevant soft
mode is either the Cooperon, or the diffuson, which are electronic
particle-hole excitations that are distinct from, but couple to, the current
mode whose correlations determine the conductivity. An example of a
non-electronic mode that couples to the current is an acoustic phonon in the
presence of an electron-phonon coupling. Since the latter is relatively weak,
this leads to a $T$-dependence that is far subleading to the Fermi-liquid $T^2$
contribution, namely, the well-known Bloch-Gr{\"u}neisen $T^5$ law.
\begin{figure}[t]
\vskip -0mm
\includegraphics[width=6.0cm]{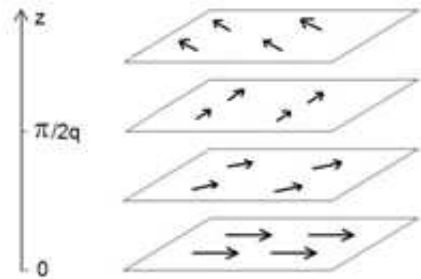}
\caption{Schematic magnetization pattern in a helimagnet.}
\label{fig:1}
\end{figure}

In this paper we consider the exchange of a more exotic soft mode, namely, the
helimagnon excitation in metallic helimagnets. Helimagnets, the best known
examples of which are MnSi and FeGe, form a class of magnetic materials with a
preferred axis in spin space, characterized by a vector ${\bm q}$. The
magnetization displays ferromagnetic order in any plane perpendicular to ${\bm
q}$, but the direction of the magnetization rotates as one moves along the
${\bm q}$-axis, forming a spiral with wavelength $2\pi/q$, where $q \equiv
\vert{\bm q}\vert$ is the pitch wave number, see Fig.\ \ref{fig:1}. A theory of
the ordered phase has been developed in Refs.\
\onlinecite{Belitz_Kirkpatrick_Rosch_2006a, Belitz_Kirkpatrick_Rosch_2006b},
which we will refer to as paper I and paper II, respectively. The magnetic
order splits the conduction band, as in the case of a ferromagnet, with a
Stoner splitting $\lambda$ that is proportional to the local magnetization. The
Goldstone mode related to the broken symmetry in spin space, the helimagnon,
has been studied in paper II. As one might expect, the helimagnon has an
anisotropic frequency-momentum relation,
\bse
\label{eqs:1.5}
\be
\omega_0({\bm k}) = \sqrt{c_z k_z^2 + c_{\perp} {\bm k}_{\perp}^4}.
\label{eq:1.5a}
\ee
The helimagnon is thus ferromagnon-like in the direction perpendicular to ${\bm
q}$, and antiferromagnon-like in the direction along ${\bm q}$. Here and in
what follows we take the pitch vector ${\bm q}$ to point in the $z$-direction,
and ${\bm k}_{\perp}$ is the component of the wave vector ${\bm k}$
perpendicular to ${\bm q}$. The elastic constants $c_z$ and $c_{\perp}$ are
given by
\bea
c_z &=& \gamma_z\,\lambda^2\,q^2/k_{\text{F}}^4,
\label{eq:1.5b}\\
c_{\perp} &=& \gamma_{\perp}\,\lambda^2/k_{\text{F}}^4.
\label{eq:1.5c}
\eea
\ese
where $\kF$ is the Fermi wave number,\cite{F_footnote} and $\gamma_z$ and
$\gamma_{\perp}$ are numbers. For the model considered in papers I and II their
values are
\be
\gamma_z = 1/36 \quad,\quad \gamma_{\perp} = 1/96.
\label{eq:1.6}
\ee
We will adopt these values for the purposes of this paper.

For later reference we note that the helical magnetization structure is a
result of the spin-orbit interaction, which is weak compared to the exchange
interaction. Consequently, the pitch wave number $q$, which is proportional to
the spin-orbit interaction, is small compared to the Fermi wave number, $q/\kF
\ll 1$.

The metallic helimagnet MnSi in particular is a very well-studied material with
many unusual properties that have been reported and discussed in detail in
Refs.\ \onlinecite{Ishikawa_et_al_1976, Pfleiderer_et_al_1997,
Pfleiderer_Julian_Lonzarich_2001} and papers I and II, among others. Here we
focus on the electrical conductivity in the ordered phase, which displays
helical order with a helix wavelength $2\pi/q \approx 180\,\AA$ below a
critical temperature $T_{\text{c}} \approx 30\,{\text K}$ at ambient pressure.
Transport measurements in the ordered phase have so far shown no significant
deviations from Fermi-liquid $T^2$-behavior. Consistent with this, a
theoretical investigation of the clean limit in paper II showed that the
leading effects of the helimagnons is a $T^{5/2}$ correction to the
Fermi-liquid behavior. Our motivation for investigating disorder corrections to
the clean behavior is two-fold: First, the residual resistance of the cleanest
MnSi samples puts them in the ballistic regime, and simple considerations
suggest very interesting behavior in that regime. Second, MnSi shows very
unusual transport behavior in the {\em paramagnetic} phase, namely, a
$T^{3/2}$-behavior of the resistivity over almost three decades in
temperature.\cite{Pfleiderer_Julian_Lonzarich_2001} The origin of this is not
understood, but it is natural to speculate that remnants of the helical order,
which are observed in the same region, have something to do with it. It thus is
prudent to first do a comprehensive study of effects of the helical order in
the ordered phase, where conditions are more clearly defined.

At a technical level, adding quenched disorder to the formalism used in papers
I and II would be hard. We thus employ an effective model that was developed in
Ref.\ \onlinecite{paper_III}, which we will refer to as paper III. Equations in
papers I-III will be referenced in the format (I.x.y),
etc.\cite{paper_II_footnote}

This paper is organized as follows. In Sec.\ \ref{sec:II} we list our most
important results for the convenience of readers who may not be interested in
the technical details. In Sec.\ \ref{sec:III} we set up a transport theory
based on the effective model, using the Kubo formalism. We first check, and
demonstrate the efficiency of, the model by reproducing the clean-limit results
of paper II in Sec. \ \ref{subsec:III.A}, and then proceed with the calculation
in the ballistic limit in Sec.\ \ref{subsec:III.B}. We discuss our results in
Sec.\ \ref{sec:IV}. Some general points pertinent to transport theory are made
in Appendix \ref{app:A}, and several calculational details are relegated to
additional appendices.

\section{Results}
\label{sec:II}

Since the details of the transport calculation for helimagnets are quite
technical, we first list our most pertinent results without any derivation.

In the helically ordered phase, the conductivity tensor is diagonal, but not
isotropic. Taking the pitch vector ${\bm q}$ in $z$-direction, its nonzero
elements are $\sigma_{zz} \equiv \sigma_{\text{L}}$, and $\sigma_{xx} =
\sigma_{yy} \equiv \sigma_{\perp}$. We find the leading correction to the
conductivity in the ballistic regime in 3D to be proportional to $\tau T$. For
a cubic crystal structure, as is the case for MnSi, we have
\be
\delta\sigma_{\text{L}} = 3\delta\sigma_{\perp} = -\sigma_0\, \frac{\pi\nu^2
\sqrt{6}}{8}\, \left(\frac{\epsilon_{\text{F}}}{\lambda}\right)^2\, \left(
\frac{q}{k_{\text{F}}}\right)^3\, \frac{T}{\epsilon_{\text{F}}}\ ,
\label{eq:2.1}
\ee
where $\nu$ is a parameter measuring deviations from a spherical Fermi surface,
see Eq.\ (\ref{eq:3.2}) below, and the prefactor of the $T$-dependence is
accurate to lowest order in the small parameter $q/\kF$.

This result is valid in a window of intermediate temperatures. For
asymptotically low temperatures, one finds diffusive, rather than ballistic,
transport behavior, and for higher temperatures the ballistic behavior crosses
over to either the Fermi-liquid behavior or the clean helimagnet conductivity,
which is proportional to $1/T^{5/2}$, see paper II. For realistic parameter
values (for known helimagnets), the temperature window is
\be
T_{\text{ball}} \ll T \ll T_{\text{ball}}\, (\epsilon_{\text{F}}\tau)\,
(q/k_{\text{F}})^3\, (\epsilon_{\text{F}}/\lambda)^3 ,
\label{eq:2.2}
\ee
where $T_{\text{ball}} \propto \lambda/(\epsilon_{\text{F}}\tau)^2$ is the
lower limit of the ballistic regime. In this regime the dominant temperature
dependence of the conductivity is given by Eq.\ (\ref{eq:2.1}), and the
conductivity is
\bea
\sigma_{\text{L},\perp} = \sigma_0 + \delta\sigma_{\text{L},\perp}(T).
\label{eq:2.3}
\eea
Notice that the sign of the correction is opposite to the Coulomb case, Eq.\
(\ref{eq:1.4a}). That is, the effect of the helimagnon exchange is {\em
antilocalizing}. We will derive these results in Sec.\ \ref{sec:III} and
discuss them in Sec.\ \ref{sec:IV}.

\section{Electrical conductivity of itinerant helimagnets}
\label{sec:III}

We now set up a standard technical formalism for transport theory in the
context of the effective model for metallic helimagnets that was given in Eq.
(III.2.19). The electrical conductivity tensor $\sigma_{ij}$ can be expressed
in terms of an equilibrium current-current correlation function by means of the
Kubo formula
\bse
\label{eqs:3.1}
\be
\sigma_{ij}(i\Omega) = \frac{i}{i\Omega}\,\left[\pi_{ij}(i\Omega) -
\pi_{ij}(i\Omega=0)\right]\ ,
\label{eq:3.1a}
\ee
where
\bea
\pi_{ij}(i\Omega) &=& -e^2 T \sum_{i\omega_1,i\omega_2} \frac{1}{V}\sum_{{\bm
p}_1,{\bm p}_{\,2}} j_i({\bm p}_1)\,j_j({\bm p}_{\, 2}) \sum_{
{\sigma_1},{\sigma_2}}
\nonumber\\
&& \hskip -30pt \times \bigl\langle{\bar\eta}_{\sigma_1}({\bm
p}_1,i\omega_1)\,\eta_{\sigma_1}({\bm p}_1,i\omega_1+i\Omega)\,
\nonumber\\
&&\hskip -0pt \times {\bar\eta}_{\sigma_2}({\bm p}_{\,2},i\omega_2)\,
        \eta_{\sigma_2}({\bm p}_{\,2},i\omega_2-i\Omega)\bigr\rangle .
\label{eq:3.1b}
\eea
\ese
is the current-current susceptibility or polarization function, with
${\bar\eta}$ and $\eta$ the fermionic fields. $\left\langle \ldots
\right\rangle$ denotes an average with respect to the action in Eq.\
(III.2.19). $i\omega \equiv i\omega_n = i2\pi T(n+1/2)$ and $i\Omega \equiv
i\Omega_n = i2\pi Tn$ $(n=\text{integer})$ are fermionic and bosonic Matsubara
frequencies, respectively, and for simplicity we suppress the index $n$.
$j({\bm p}) =
\partial\epsilon_{\bm p}/\partial {\bm p}$ is the current vertex, and for the
electronic energy-momentum relation we use an expression appropriate for a
cubic crystal structure,
\be
\epsilon_{\bm p} = \frac{{\bm p}^{\,2}}{2\me} + \frac{\nu}{2m_{\text{e}}
k_{\text{F}}^2}\, \left(p_x^{\,2} p_y^{\,2} + p_y^{\,2} p_z^{\,2} + p_z^{\,2}
p_x^{\,2} \right) + O(p^{\,6}),
\label{eq:3.2}
\ee
with $\me$ the electronic effective mass, and $\nu = O(1)$ a dimensionless
coupling constant that measures deviations from a spherical Fermi surface.

The conductivity as written above is actually the transport coefficient for the
quasiparticles defined in paper III, which are described by the fermionic
fields ${\bar\eta}$ and $\eta$. The physical conductivity is given in terms of
the electron fields ${\bar\psi}$ and $\psi$, which are related to the
quasiparticle fields by the transformation given in Eqs. (III.2.9). However, we
will work to lowest order in the small parameter $q/\kF$, and to this accuracy
the quasiparticle conductivity is the same as the physical conductivity, as can
readily be seen from Eqs.\ (III.2.9).

The four-point fermionic correlation function in Eq.\ (\ref{eq:3.1b}) is
conveniently expressed in terms of Green functions ${\cal G}$ and a vector
vertex function ${\bm\Gamma}$ with components $\Gamma^{\,i}$,
\begin{figure}[t]
\vskip -0mm
\includegraphics[width=6.0cm]{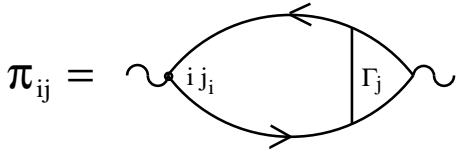}
\caption{Graphic representation of the polarization function. The directed
 solid lines denote Green functions.}
\label{fig:2}
\end{figure}
\begin{figure}[t]
\vskip -0mm
\includegraphics[width=8.0cm]{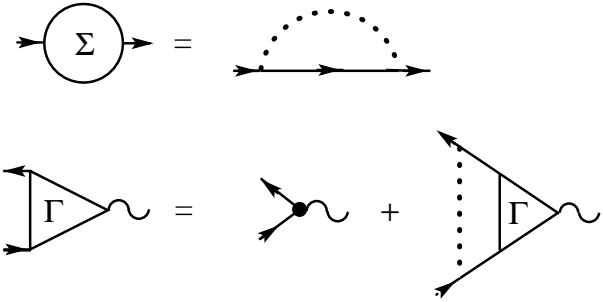}
\caption{Conserving approximation for the self energy and the vertex function.
 The dotted line denotes the effective potential.}
\label{fig:3}
\end{figure}
\bea
\pi_{ij}(i\Omega) &=& -e^2 T \sum_{i\omega}\frac{1}{V} \sum_{\sigma} i
j_{\,i}({\bm p})\, {\cal G}_{\sigma}({\bm p},i\omega)\, {\cal G}_{\sigma}({\bm
p},i\omega - i\Omega)
\nonumber\\
&&\hskip 50pt \times \Gamma^j_{\sigma}({\bm p}\,;i\omega,i\omega - i\Omega),
\label{eq:3.3}
\eea
see Fig.\ \ref{fig:2}. This expression is valid if the Green function
$\langle\eta_{\sigma_1}({\bm p}_1,i\omega_1)\,{\bar\eta}_{\sigma_2}({\bm
p}_{\,2},i\omega_2)\rangle$ is diagonal in both momentum and spin. For our
effective model this is the case (whereas it was not the case in paper II), and
${\cal G}$ is expressed in terms of the self energy $\Sigma$ by means of the
usual Dyson equation
\be
{\cal G}_{\sigma}({\bm p},i\omega) = \frac{1}{G_{0,\sigma}^{-1}({\bm
p},i\omega) - \Sigma_{\sigma}({\bm p},i\omega)}\ .
\label{eq:3.4}
\ee
Here $G_0$ is the bare Green function, which is given by Eq.\ (III.2.10c).
Equations (\ref{eq:3.3}, \ref{eq:3.4}) just shift the problem into the
determination of the self energy $\Sigma$ and the vertex function
${\bm\Gamma}$. In order to evaluate the Kubo formula, it is most convenient to
separately treat the cases with and without quenched disorder, respectively.

\subsection{Clean limit}
\label{subsec:III.A}

We first use the formalism developed so far to re-derive some of the results of
paper II for the conductivity of helimagnets in the absence of any elastic
scattering. This serves both as a check, and as a demonstration of how much
simpler it is to evaluate the Kubo formula within the quasiparticle model
compared to the model used in paper II. The calculation presented here is just
a slight generalization of what is presented in Appendix \ref{app:A}, which
serves to discuss the extent to which the approximations used are controlled.

\subsubsection{Conserving approximation for the conductivity}
\label{subsubsec:III.A.1}

It is well known that, in clean systems, care must be taken to treat the self
energy $\Sigma$, which enters the Green function $\cal{G}$, and the vertex
function ${\bm\Gamma}$ consistently in a conserving
approximation.\cite{Baym_Kadanoff_1961, Kadanoff_Baym_1962} The simplest
consistent approximation, which is equivalent to the Boltzmann equation, is to
treat the self energy in a self-consistent Born approximation, and the vertex
function in a ladder approximation. These are graphically represented in Fig.\
\ref{fig:3}. Analytically we have integral equations
\bse
\label{eqs:3.5}
\bea
\Sigma_{\sigma}({\bm p},i\omega) &=& -T\sum_{i\Omega} \frac{1}{V}\sum_{{\bm k}}
V({\bm k};{\bm p}-{\bm k},{\bm p};i\Omega) \nonumber\\
&& \times {\cal G}_{\sigma}({\bm p}-{\bm k},i\omega - i\Omega),
\label{eq:3.5a}
\eea
for the self energy, and
\begin{widetext}
\bea
{\bm\Gamma}_{\sigma}({\bm p};i\omega,i\omega - i\Omega) &=& i{\bm j}({\bm p}) -
\frac{1}{V}\sum_{\bm k} T\sum_{i\Omega'} V({\bm k};{\bm p}-{\bm k},{\bm
p};i\Omega')\, {\cal G}_{\sigma}({\bm p}-{\bm k},i\omega - i\Omega')\, {\cal
G}_{\sigma}({\bm p}-{\bm k},i\omega - i\Omega' - i\Omega)\,
\nonumber\\
&& \hskip 150pt \times {\bm\Gamma}_{\sigma}({\bm p}-{\bm k};i\omega -
i\Omega',i\omega - i\Omega' - i\Omega),
\label{eq:3.5b}
\eea
\end{widetext}
\ese
for the vertex function. $V$ is the effective potential from Eqs.\ (III.2.18).

For completeness, we list the bare Green function from Eq. (III.2.10c),
\bse
\label{eqs:3.6}
\be
G_{0,\sigma}({\bm p},i\omega) = \frac{1}{i\omega - \omega_{\sigma}({\bm p})}\ ,
\label{eq:3.6a}
\ee
where
\be
\omega_{\sigma}({\bm p}) = \frac{1}{2} \left(\xi_{\bm p} + \xi_{{\bm p} + {\bm
q}} - (-)^{\sigma} \sqrt{(\xi_{\bm p} - \xi_{{\bm p} + {\bm q}})^2 +
4\lambda^2}\right),
\label{eq:3.6b}
\ee
\ese
with $\xi_{\bm p} = \epsilon_{\bm p} - \epsilon_{\text{F}}$, $\sigma = 1,2$,
and the effective potential from Eqs.\ (III.2.18),\cite{screening_footnote}
\bse
\label{eqs:3.7}
\be
V({\bm k};{\bm p},{\bm p}';i\Omega) = V_0\, \chi({\bm k},i\Omega)\,\gamma({\bm
k},{\bm p})\, \gamma(-{\bm k},{\bm p}').
\label{eq:3.7a}
\ee
Here
\be
V_0 = \lambda^2 (q^2/8m_{\text{e}}^2),
\label{eq:3.7b}
\ee
and
\be
\chi({\bm k},i\Omega) = \frac{1}{2N_{\text{F}}}\, \frac{q^2/3
k_{\text{F}}^2}{\omega_0^2({\bm k}) - (i\Omega)^2}\ ,
\label{eq:3.7c}
\ee
is the helimagnon susceptibility. For later reference, we also list its
spectral function
\bea
\chi''({\bm k},u) &=& \text{Im}\,\chi({\bm k},i\Omega \rightarrow u+i0)
\nonumber\\
&& \hskip -50pt = \frac{\pi}{12 N_{\text{F}}}\, \frac{q^2}{k_{\text{F}}^2}\,
\frac{1}{\omega_0({\bm k})}\, \left[\delta(u-\omega_0({\bm k})) -
\delta(u+\omega_0({\bm k}))\right].
\nonumber\\
\label{eq:3.7d}
\eea
Finally,
\be
\gamma({\bm k},{\bm p}) = \frac{1}{2\lambda}\, \left[k_z +
\frac{\nu}{k_{\text{F}}^2} \left(k_z p_{\perp}^2 + 2({\bm k}_{\perp}\cdot{\bm
p}_{\perp}) p_z\right) \right]
\label{eq:3.7e}
\ee
\ese
is a vertex function. This specifies all input parameters for the two coupled
integral equations (\ref{eqs:3.5}).

Since the two spin projections do not couple, we can restrict ourselves to one
spin projection at a time, which effectively reduces the problem to one of
spinless electrons. In what follows, we consider the contribution from the pole
$\omega_1({\bm p})$ and drop the spin label elsewhere. In the end, the
contribution from the pole $\omega_2({\bm p})$ can simply be added.

\subsubsection{Solution of the integral equations}
\label{subsubsec:III.A.2}

The coupled integral equations (\ref{eqs:3.5}) can now be solved by following a
slight generalization of the procedure outlined in Appendix \ref{app:A}. The
conductivity is still given by Eq.\ (\ref{eq:A.5a}), and Eqs.\ (\ref{eqs:A.8})
remain valid. We thus have the single-particle relaxation rate given by
\bse
\label{eqs:3.8}
\be
\Gamma_0(\epsilon) = -\NF \int_{-\infty}^{\infty} du\
\left[n_{\text{B}}\left(\frac{u}{T}\right) +
n_{\text{F}}\left(\frac{u+\epsilon}{T}\right)\right]\, {\bar V}_0''(u),
\label{eq:3.8a}
\ee
with the zeroth moment of the potential spectrum given by
\be
{\bar V}_0''(u) = \frac{1}{\NF^2}\, \frac{1}{V^2} \sum_{{\bm k},{\bm p}}
\delta(\omega_1({\bm k}))\, \delta(\omega_1({\bm p}))\, V''({\bm k}-{\bm
p};{\bm k},{\bm p};u),
\label{eq:3.8b}
\ee
\ese
with the potential $V$ from Eqs.\ (\ref{eqs:3.7}). This leads to
\bse
\label{eqs:3.9}
\begin{widetext}
\bea
\Gamma_0({\epsilon}) &=& \frac{-\nu^2 q^2}{8\me^2\kF^4\NF}
\int_{-\infty}^{\infty} du\,\left[n_{\text{B}}\left(\frac{u}{T}\right) +
n_{\text{F}}\left(\frac{u+\epsilon}{T}\right)\right]\ \frac{1}{V^2} \sum_{{\bm
k},{\bm p}} \delta(\omega_1({\bm k}))\, \delta(\omega_1({\bm p}))\, k_z\, p_z\,
[{\bm k}_{\perp}\cdot({\bm p}_{\perp}-{\bm k}_{\perp})]\,
\nonumber\\
&&\hskip 250pt \times [{\bm p}_{\perp}\cdot({\bm k}_{\perp}-{\bm
p}_{\perp})]\,\chi''({\bm k}-{\bm p},u).
\label{eq:3.9a}
\eea
Here $n_{\text{B}}(x) = 1/(e^x-1)$ and $n_{\text{F}}(x) = 1/(e^x+1)$ are the
Bose and Fermi distribution functions, respectively. Since the susceptibility
$\chi$ is soft at zero wave number, to leading order in the temperature this
can be rewritten as
\be
\Gamma_0({\epsilon}) = \frac{\nu^2 q^2}{8\me^2\kF^4\NF} \int_{-\infty}^{\infty}
du\,\left[n_{\text{B}}\left(\frac{u}{T}\right) +
n_{\text{F}}\left(\frac{u+\epsilon}{T}\right)\right]\ \frac{1}{V^2} \sum_{{\bm
k},{\bm p}} \delta(\omega_1({\bm k}))\, \delta(\omega_1({\bm p}))\,
k_z^2\,\left[{\bm k}_{\perp}\cdot({\bm p}_{\perp}-{\bm k}_{\perp})\right]^2\,
\chi''({\bm k}-{\bm p},u).
\label{eq:3.9b}
\ee
The same result is obtained from Eq.\ (III.3.5) by averaging $1/\tau({\bm
p},\epsilon)$ over the Fermi surface. Evaluating the integral leads to
\be
\Gamma_0(\epsilon) = \frac{5\pi\, \nu^4\,g(\nu)}{1,024\times 6^{3/4}}\ \lambda
\left(\frac{q}{\kF}\right)^6 \left(\frac{\epsilon_{\text{F}}}{\lambda}\right)^2
\left(\frac{T}{T_q}\right)^{3/2} \gamma_0(\epsilon/2T),
\label{eq:3.9c}
\ee
with $T_q = \lambda q^2/6\kF^2$. Here
\bea
g(\nu) = \frac{16^3}{5\pi^2} \int_0^1 d\eta\ \eta^2(1-\eta^2)^{5/2}
\int_0^{2\pi} d\varphi\ \frac{\sin^2\varphi\, \cos^2\varphi\,
\cos^2(2\varphi)}{[D(\nu,\eta,\varphi)]^{3/2}}\ , \nonumber\\
\label{eq:3.9d}
\eea
with
\be
D(\nu,\eta,\varphi) = 1+2\nu[\eta^2 + 2(1-\eta^2) \sin^2\varphi\,
\cos^2\varphi] + \nu^2[\eta^4+(1+2\eta^2-3\eta^4) \sin^2\varphi \cos^2\varphi]\
.
\label{eq:3.9e}
\ee
\end{widetext}
We have normalized $g$ such that $g(\nu=0)=1$. $\gamma_0$ is the $n=0$ member
of a family of functions defined by
\bea
\gamma_n(y) &=& \frac{K_{(2n+1)/2}}{32} \int_0^{\infty} dr\, r^{(2n+1)/2}\
\bigl[2n_{\text{B}}(r)
\nonumber\\
&&\hskip 25pt + n_{\text{F}}(r+2y) + n_{\text{F}}(r-2y)\bigr],
\label{eq:3.9f}
\eea
with
\be
K_{\mu} = 4\int_0^{\pi/2}dx\ \sin^{\mu}x =
2^{\mu+1}\Gamma^{\,2}\left(\frac{\mu+1}{2}\right)/\Gamma(\mu+1)\,,
\label{eq:3.9g}
\ee
\ese
where $\Gamma$ denotes the Gamma function. The same result is obtained by
integrating Eq.\ (II.3.29d) or (III.3.6a) over the Fermi surface. For an
explanation of the physical relevance of the temperature scale $T_q$, see the
discussion after Eq.\ (\ref{eq:4.5}) below.

Similarly, Eqs.\ (\ref{eqs:A.8}) still hold, and we find a transport relaxation
rate
\bse
\label{eqs:3.10}
\be
\Gamma_1(\epsilon) = -\NF \int_{-\infty}^{\infty} du\
\left[n_{\text{B}}\left(\frac{u}{T}\right) +
n_{\text{F}}\left(\frac{u+\epsilon}{T}\right)\right]\, {\bar V}_1''(u),
\label{eq:3.10a}
\ee
with
\bea
{\bar V}_1''(u) &=& \frac{1}{\NF^2}\, \frac{1}{V^2} \sum_{{\bm k},{\bm p}}
\delta(\omega_1({\bm k}))\, \delta(\omega_1({\bm p}))\, \frac{({\bm k}-{\bm
p})^2}{2\kF^2}\, \nonumber\\
&&\hskip 50pt \times V''({\bm k}-{\bm p};{\bm k},{\bm p};u).
\label{eq:3.10b}
\eea
This agrees with Eq.\ (II.3.39b) after integration over the Fermi surface.
Explicitly, we find
\be
\Gamma_1(\epsilon) = \frac{5\pi\, \nu^4\,g(\nu)}{512\times 6^{5/4}}\ \lambda
\left(\frac{q}{\kF}\right)^8 \left(\frac{\epsilon_{\text{F}}}{\lambda}\right)^2
\left(\frac{T}{T_q}\right)^{5/2} \gamma_1(\epsilon/2T).
\label{eq:3.10c}
\ee
\ese
Here $\gamma_1$ is given by Eq.\ (\ref{eq:3.9f}) with $n=1$. The conductivity
is given by
\bse
\label{eqs:3.11}
\be
\sigma = \frac{e^2 \kF^2 \NF}{6\me^2} \int_{-\infty}^{\infty}
\frac{d\epsilon}{4T}\,\frac{1}{\cosh^2(\epsilon/2T)}\,
\frac{1}{\Gamma_1({\epsilon})}\ ,
\label{eq:3.11a}
\ee
which leads to a Drude formula
\be
\sigma = ne^2\tau_1/\me.
\label{eq:3.11b}
\ee
Here $n = \kF^3/6\pi^2$ is the electron density per spin, since we are
considering an effectively spinless problem, see the remark at the end of
\ref{subsubsec:III.A.1}. The transport relaxation rate is
\be
1/\tau_1 = \frac{\nu^4 g(\nu)}{C_1}\,\lambda\, \left(\frac{q}{\kF}\right)^8
\left(\frac{\epsilon_{\text{F}}}{\lambda}\right)^2
\left(\frac{T}{T_q}\right)^{5/2},
\label{eq:3.11c}
\ee
in agreement with Eq.\ (II.3.40a). Here
\be
C_1 = \frac{256\times 6^{5/4}}{5\pi} \int_{0}^{\infty}
\frac{dx}{\cosh^2(x)\,\gamma_1(x)} \approx 186\ .
\label{eq:3.11d}
\ee
\ese
In deriving Eqs.\ (\ref{eq:3.10b}) - (\ref{eq:3.11d}) we have assumed a
spherical Fermi surface whenever doing so does not make the integral vanish. As
a consequence, the $\nu$-dependence of the prefactors in Eqs.\ (\ref{eq:3.10c})
and (\ref{eq:3.11d}) is exact, within our model, only to lowest order in $\nu$.

We see that the current effective model reproduces the results of paper II for
the clean case, and we have also calculated some of the prefactors that were
not given explicitly in paper II.

\subsection{Ballistic limit}
\label{subsec:III.B}

We now add quenched disorder to the action, using the standard impurity model
with an elastic relaxation time $\tau$ described in paper III. We then need to
distinguish between the diffusive limit, where the relaxation rate $1/\tau$ is
large compared to the frequency or temperature in appropriate units, and the
ballistic one, where the opposite inequality holds. In the diffusive limit, it
is well known that an infinite resummation of impurity diagrams is needed to
work to a given order in the disorder. In the ballistic limit, this is not the
case, and a straightforward diagrammatic perturbative expansion in the number
of impurity lines is possible. This yields impurity corrections to the clean
conductivity. For the case of electrons interacting via a screened Coulomb
interaction, this has been investigated by Zala et al.,
\cite{Zala_Narozhny_Aleiner_2001} and the development in the present case
follows the same general lines.
\begin{figure}[b]
\vskip -0mm
\includegraphics[width=8.0cm]{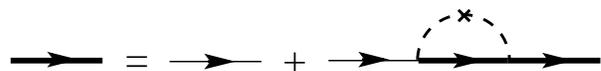}
\caption{Defining equation for the Green function $G$ (thick solid lines) in
 terms of the bare Green function $G_0$ (thin solid lines) and the impurity
 factor $u_0$ (dashed line with cross).}
\label{fig:4}
\end{figure}

It is convenient to include the elastic relaxation rate in the bare Green
function in a self-consistent Born approximation, see Fig.\ \ref{fig:4}. That
is, instead of the bare Green function $G_0$, Eq.\ (\ref{eq:3.6a}), we use
\be
G({\bm p},i\omega) = \frac{1}{i\omega - \omega_1({\bm p}) +
(i/2\tau)\sgn\omega}\ .
\label{eq:3.12}
\ee
Here we absorb the correction to the bare elastic relaxation rate that was
discussed in Sec. III.A of paper III in $1/\tau$. In addition to using $G$
instead of $G_0$, diagrams must be decorated with explicit impurity lines,
which diagrammatically are denoted by dashed lines with crosses, and which
carry a factor
\be
u_0 = 1/2\pi N_{\text{F}}\tau.
\label{eq:3.13}
\ee
The Green function ${\cal G}$, Eq.\ (\ref{eq:3.4}), can now be written
\be
{\cal G}({\bm p},i\omega) = \frac{1}{G^{-1}({\bm p},i\omega) -
\delta\Sigma({\bm p},i\omega)}\ ,
\label{eq:3.14}
\ee
where the self energy $\delta\Sigma$ does not contain the simple impurity
self-energy that is incorporated in $G$.

\begin{figure*}[t,h,b]
\includegraphics[width=16.0cm]{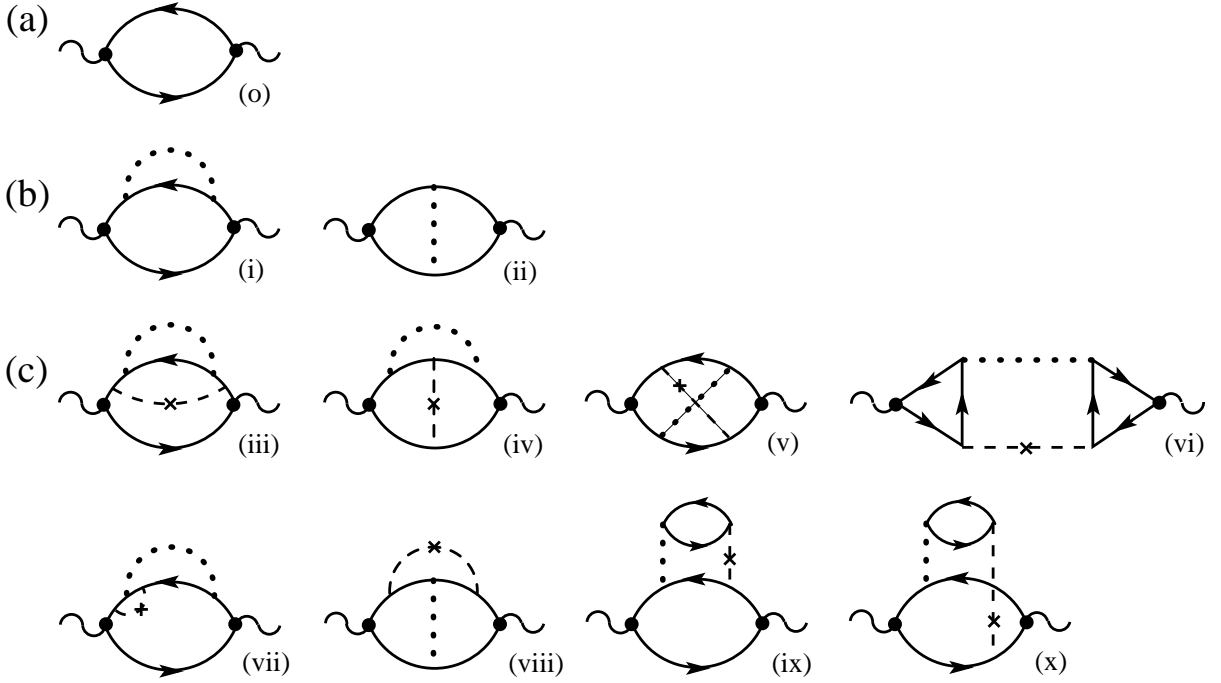}
\caption{Leading disorder corrections to the clean conductivity. Solid lines
 denote the Green function $G$, dotted lines denote the effective potential,
 and dashed lines with crosses denote the impurity factor $u_0$.}
\label{fig:5}
\end{figure*}

We are interested in the leading disorder correction to the clean resistivity
calculated in paper II, and in the leading temperature dependence of that
correction. To find this, it suffices to work to first order in both the
disorder and the effective potential,\cite{perturbation_footnote} and we can
expand the conductivity up to linear order in $\delta\Sigma$ and the vertex
function ${\bm\Gamma}$. From Eqs.\ (\ref{eq:3.1a}, \ref{eq:3.3}) and
(\ref{eq:3.14}) we find the following expression for the static conductivity
$\sigma_{ij} = \text{Re} \lim_{\Omega\rightarrow 0} \sigma_{ij}(i\Omega
\rightarrow \Omega i0)$:
\bse
\label{eqs:3.15}
\be
\sigma_{ij} = \sigma_{ij}^{(0)} + \delta\sigma_{ij}^{\Sigma} +
\delta\sigma_{ij}^{\Gamma},
\label{eq:3.15a}
\ee
with
\begin{widetext}
\bea
\sigma_{ij}^{(0)} &=& \frac{1}{V}\sum_{\bm p} j_i({\bm p})\,j_j({\bm p})\,
\frac{1}{2T} \int \frac{d\epsilon}{4\pi}\, \frac{1}{\cosh^2(\epsilon/2T)}\
\left[G_R({\bm p},\epsilon)\,G_A({\bm p},\epsilon) - \text{Re}\, \left(G_R({\bm
p},\epsilon)\right)^2\right],
\label{eq:3.15b}\\
\delta\sigma_{ij}^{\Sigma} &=& \frac{1}{V}\sum_{\bm p} j_i({\bm p})\,j_j({\bm
p})\, \frac{1}{2T} \int \frac{d\epsilon}{4\pi}\,
\frac{1}{\cosh^2(\epsilon/2T)}\ 2\text{Re}\,\left[\left(G_R({\bm
p},\epsilon)\right)^2\,G_A({\bm p},\epsilon)\,\delta\Sigma_R({\bm p},\epsilon)
+ \left(G_R({\bm p},\epsilon)\right)^3\,\delta\Sigma_R({\bm p},\epsilon)
\right], \nonumber\\
\label{eq:3.15c}\\
\delta\sigma_{ij}^{\Gamma} &=& \frac{1}{V}\sum_{\bm p} j_i({\bm p})\,
\frac{1}{2T} \int \frac{d\epsilon}{4\pi}\, \frac{1}{\cosh^2(\epsilon/2T)}\
\text{Re}\,\Bigl[ G_R({\bm p},\epsilon)\, G_A({\bm p},\epsilon)\, \Gamma_j({\bm
p};\epsilon+i0,\epsilon-i0)
\nonumber\\
&&\hskip 250pt - \left(G_R({\bm p},\epsilon)\right)^2\,
\Gamma_j({\bm p};\epsilon+i0,\epsilon+i0)\Bigr]. \nonumber\\
\label{eq:3.15d}
\eea
\end{widetext}
\ese
To write Eqs.\ (\ref{eqs:3.15}) we have performed the Matsubara frequency sums
and have introduced retarded and advanced Green functions $G_{R,A}({\bm
p},\epsilon) = G({\bm p},i\omega \rightarrow \epsilon \pm i0)$, and a retarded
self energy $\delta\Sigma_R({\bm p},\epsilon) = \delta\Sigma({\bm p},i\omega
\rightarrow \epsilon + i0)$. Diagrammatically, these contributions to the
conductivity are shown in Fig.\ \ref{fig:5}.
In evaluating these diagrams, we again make use of the small parameter
$q/k_{\text{F}} \ll 1$. To lowest order in $q/k_{\text{F}}$, in many cases the
Green function $G$ can be replaced by the free-electron Green function, which
greatly simplifies the integrals.

We further notice that the conductivity tensor is not isotropic, since the
integrand depends on the helix pitch vector ${\bm q}$. However, simple symmetry
considerations show that it is still diagonal, with different components in the
directions parallel and perpendicular to ${\bm q}$, respectively,
\be
\delta\sigma_{ij} = \delta_{ij}\,\left[\delta_{iz}\,\delta\sigma_{\text{L}} +
(1-\delta_{iz})\,\delta\sigma_{\perp}\right].
\label{eq:3.16}
\ee

The diagrams can be classified as follows. Diagram (o) in Fig.\ \ref{fig:5}(a)
represents $\sigma^{(0)}$. To lowest order in the disorder, and in
$q/k_{\text{F}}$, it yields the Drude conductivity, Eq.\ (\ref{eq:1.1}),
\be
\sigma_{ij}^{(0)} = \delta_{ij}\,\sigma_0\,\left[1 +
O\left(1/\epsilon_{\text{F}}\tau, (q/k_{\text{F}})^2\right)\right].
\label{eq:3.17}
\ee

Diagrams (i), (iii), (vii), and (ix) contribute to $\delta\sigma^{\Sigma}$, and
the remaining diagrams contribute to $\delta\sigma^{\Gamma}$. Diagrams (i) and
(ii) in Fig.\ \ref{fig:5}(b) do not contain explicit impurity lines, and hence
need to be evaluated to next-to-leading order in the disorder. The diagrams in
Fig.\ \ref{fig:5}(c) contain an explicit impurity line, and evaluating them to
leading order suffices.

\subsubsection{Diagrams without explicit impurity lines}
\label{subsubsec:III.B.1}

Let us first consider the diagrams (i) and (ii). Standard techniques yield
\begin{widetext}
\bse
\label{eqs:3.18}
\bea
\delta\sigma^{(i)}_{ij} &=& \frac{-V_0}{4\pi}\,\frac{1}{T}
\int_{-\infty}^{\infty} \frac{d\epsilon}{\cosh^2(\epsilon/2T)}\,\frac{1}{V}
\sum_{\bm k} \int_{-\infty}^{\infty} \frac{du}{\pi}\, \chi''({\bm k},u)\
\Biggl[n_{\text{B}}\left(\frac{u}{T}\right)\,\text{Re}\,J_{ij}^{++-,+}(k) \nonumber\\
&& \hskip 160pt + \frac{1}{2}\, n_{\text{F}}\left(\frac{u-\epsilon}{T}\right)
\, \text{Re}\, \left[J_{ij}^{++-,+}(k) - J_{ij}^{++-,-}(k)\right]\Biggr],
\label{eq:3.18a}\\
\delta\sigma^{(ii)}_{ij} &=& \frac{-V_0}{8\pi}\,\frac{1}{T}
\int_{-\infty}^{\infty} \frac{d\epsilon}{\cosh^2(\epsilon/2T)}\,\frac{1}{V}
\sum_{\bm k} \int_{-\infty}^{\infty} \frac{du}{\pi}\, \chi''({\bm k},u)\
\Biggl[n_{\text{B}}\left(\frac{u}{T}\right)\,\text{Re}\, \left[
J_{ij}^{+-,+-}(k) - I_{ij}^{+-,+-}(k)\right] \nonumber\\
&& \hskip 60pt + n_{\text{F}}\left(\frac{u-\epsilon}{T}\right) \, \text{Re}\,
\left[ J_{ij}^{+-,+-}(k) - I_{ij}^{+-,+-}(k) - J_{ij}^{+-,++}(k) +
I_{ij}^{+-,++}(k)\right] \Biggr].
\label{eq:3.18b}
\eea
\ese
Here the $J$ are defined by convolutions of Green functions,
\bse
\label{eqs:3.19}
\bea
J_{ij}^{++-,+}(k) &=& \frac{1}{V}\sum_{\bm p}\, j_i({\bm p})\, j_j({\bm
p})\,\gamma({\bm k},{\bm p}) \, \gamma({\bm k},{\bm p}-{\bm k})\, G_R({\bm
p})\, G_R({\bm p}) \,G_A({\bm p}) \, G_R({\bm p}-{\bm k}),
\label{eq:3.19a}\\
J_{ij}^{++-,-}(k) &=& \frac{1}{V}\sum_{\bm p}\, j_i({\bm p})\, j_j({\bm
p})\,\gamma({\bm k},{\bm p}) \, \gamma({\bm k},{\bm p}-{\bm k})\, G_R({\bm
p})\, G_R({\bm p}) \,G_A({\bm p}) \, G_A({\bm p}-{\bm k}),
\label{eq:3.19b}\\
J_{ij}^{+-,+-}(k) &=& \frac{1}{V}\sum_{\bm p}\, j_i({\bm p})\, j_j({\bm
p})\,\gamma({\bm k},{\bm p}) \, \gamma({\bm k},{\bm p}-{\bm k})\, G_R({\bm
p})\,G_A({\bm p}) \, G_R({\bm p}-{\bm k})\, G_A({\bm p}-{\bm k}),
\label{eq:3.19c}\\
I_{ij}^{+-,+-}(k) &=& \frac{1}{V}\sum_{\bm p}\, j_i({\bm p})\, j_j({\bm
k})\,\gamma({\bm k},{\bm p}) \, \gamma({\bm k},{\bm p}-{\bm k})\, G_R({\bm
p})\,G_A({\bm p}) \, G_R({\bm p}-{\bm k})\, G_A({\bm p}-{\bm k}),
\label{eq:3.19d}
\eea
\ese
\end{widetext}
where $G_{R,A}({\bm p}) = G_{R,A}({\bm p},\epsilon=0)$. Other convolutions are
defined analogously, with the upper $\pm$ indices denoting retarded and
advanced Green functions, and the comma separating them denoting the momentum
structure of the convolution. In writing Eqs.\ (\ref{eqs:3.18}) we have
neglected contributions from other convolutions of four Green functions that
are easily shown to be of higher order in the disorder than the ones we kept.
For instance, a complete expression for diagram (i) contains contributions from
$J_{ij}^{+++,-}$ and $J_{ij}^{+++,+}$, which are subleading in this sense.
Also, a complete evaluation of the diagrams yields nominal contributions
proportional to $\chi'$, the Kramers-Kronig transform of $\chi''$. These vanish
once the real part is taken, as is to be expected: by Fermi's golden rule, to
first order in the interaction potential, the scattering cross-section and
hence the conductivity depend only on the spectrum of the potential. Finally,
we have used the fact that the internal frequencies $u$ and $\epsilon$ in Eqs.\
(\ref{eqs:3.18}) scale as the temperature $T$. To find the leading temperature
dependence, we therefore can drop the frequency dependence of the Green
functions, and this is reflected in Eqs.\ (\ref{eqs:3.19}).

To evaluate the integrals in Eqs.\ (\ref{eqs:3.19}) we work to lowest order in
$q/k_{\text{F}}$. We further neglect $\lambda$, since in our effective
single-spin-projection model it amounts (at $q=0$) to just a shift of the Fermi
energy. That is, we replace $\omega_1({\bm p})$ in Eq. (\ref{eq:3.8b}) by
$\xi_{\bm p}$. We further use a nearly-free electron expression for
$\xi_{\bm{p}}$, i.e., we put $\nu=0$ in Eq. (\ref{eq:3.2}). By comparison with
paper II, we see that this does not qualitatively affect the result, see below.
These simplifications lead in particular to $j_i({\bm p}) = p_i/m_{\text{e}}$,
and to lowest order in the disorder the integrals can be evaluated in the
familiar approximation that replaces the integration over $\vert{\bm p}\vert$
by a contour integration over $\xi_{\bm
p}$,\cite{Abrikosov_Gorkov_Dzyaloshinski_1963} which we will refer to as the
AGD approximation. Power counting shows that the leading individual
contributions to $\delta\sigma$ are of order $\delta\sigma \propto \tau^2
T^{3/2}$. This should be understood as the second term in an expansion of
$1/(1/\tau + T^{3/2})$. (We recall that the clean single-particle relaxation
rate is proportional to $T^{3/2}$.) We know from paper II, and from Sec.\
\ref{subsec:III.A} above, that these terms must cancel, and that the leading
temperature dependence at $O(\tau^2)$ is $\tau^2 T^{5/2}$. These terms, and
higher ones in the diverging disorder expansion, must be resummed to yield
$1/(1/\tau + T^{5/2})$, which is the inverse of the clean-limit transport rate
added to the elastic scattering rate according to Matthiessen's rule.
Corrections to these contributions are smaller by a factor of $1/\tau T^{1/2}$,
which leads to conductivity corrections $\delta\sigma \propto \tau T$. These
all cancel by the same mechanism that leads to the cancellation of the $\tau^2
T^{3/2}$ terms, and this can be seen without performing the integrals. Finally,
the convolutions $I$ in Eq.\ (\ref{eq:3.18b}) are subleading in temperature
compared to the $J$ by power counting: at $O(\tau^2)$ they contribute to the
clean-limit $T^{5/2}$ term, and the leading corrections are again small by a
factor of $1/\tau T^{1/2}$, which leads to conductivity corrections
$\delta\sigma \propto \tau T^2$. We thus obtain the following result:
\be
\delta\sigma_{ij}^{(i)} + \delta\sigma_{ij}^{(ii)} = O(\tau^2 T^{5/2}) + o(\tau
T),
\label{eq:3.20}
\ee
where $o(x)$ denotes terms that are smaller than $O(x)$. The arguments leading
to this conclusion are outlined in Appendix \ref{app:B}. The term of
$O(\tau^2)$ was interpreted above, and we will not calculate the leading
temperature dependence of the term of $O(\tau)$ since we will find
contributions of $O(\tau T)$ from other diagrams.

\subsubsection{Diagrams with explicit impurity lines}
\label{subsubsec:III.B.2}

We now turn to the diagrams in Fig.\ \ref{fig:5}(c), which carry an explicit
impurity line. Their contribution to the conductivity is of $O(\tau)$, and it
thus suffices to calculate them to leading order in the disorder. Before we do
so, we identify the small parameter that controls our disorder expansion. As we
point out in Appendix \ref{app:B}, the expansion parameter for the convolutions
$J$ that appear in the integrand in Eqs.\ (\ref{eqs:3.18}) is $\delta = 1/\vF
k_{\perp}\tau$, with $\vF = \kF/\me$ the Fermi velocity. According to Eqs.\
(\ref{eqs:1.5}), the transverse wave number scales as the square root of the
helimagnon frequency, which in turn scales as the temperature by virtue of
Eqs.\ (\ref{eq:3.7c}) and (\ref{eqs:3.18}). The small expansion parameter is
thus
\be
\delta = 1/4\times 6^{1/4}\sqrt{(\epsilon_{\text{F}}\tau)^2 T/\lambda}\ ,
\label{eq:3.21}
\ee
and this will turn out to be true for the diagrams in Fig.\ \ref{fig:5}(c) as
well. This is different from the Coulomb case, where the small parameter that
controls the ballistic regime is $1/T\tau$,\cite{Zala_Narozhny_Aleiner_2001}
and it will be important for discussing the size of the ballistic regime in
Sec.\ \ref{sec:IV} below.

The diagrams in Fig.\ \ref{fig:5}(c) all contain six Green functions that
factorize into two sets of momentum convolutions containing $n$ and $6-n$ Green
functions, respectively, with $n=3$ or $n=4$. Diagrams (iii) - (vi) contain the
$(3,3)$ partitions, whereas diagrams (vii) - (x) contain the $(4,2)$
partitions. The same power-counting arguments that we employed for diagrams (i)
and (ii), and that are explained in Appendix \ref{app:B}, reveal the following:

\smallskip\noindent First, to lowest order in the small parameter
$(q/k_{\text{F}})$ (i.e., replacing the helimagnon Green functions by
nearly-free electron Green functions), only the $(3,3)$ partitions contribute
to $O(\tau T)$, whereas the $(4,2)$ partitions are of higher order in the
temperature. That is,
\be
\delta\sigma_{ij}^{(vii)-(x)} = o(\tau T)
\label{eq:3.22}
\ee
to lowest order in $(q/k_{\text{F}})$, and we will evaluate all other diagrams
to lowest order in this small parameter as well. We will come back to what
happens to higher order in $(q/k_{\text{F}})$ in Sec.\ \ref{subsubsec:III.B.3}
below.

\smallskip\noindent
Second, for the transverse conductivity correction $\delta\sigma_{\perp}$ only
diagram (iii) contributes to $O(\tau T)$, whereas for $\delta\sigma_{\text{L}}$
the other $(3,3)$ partitions also contribute.

\smallskip\noindent
In addition, by considering the reality properties of the convolutions
involved, one finds that, third, diagram (vi) is given in terms of the real
part of a convolution that is purely imaginary, and hence does not contribute.
Therefore, in order to obtain the transverse conductivity correction
$\delta\sigma_{\perp}$ to leading order in the small parameter $\delta$, Eq.\
(\ref{eq:3.21}), one needs to calculate only diagram (iii). For the
longitudinal correction $\delta\sigma_{\text{L}}$ one needs to also consider
diagrams (iv) and (v).

\smallskip\noindent
Finally, a cursory inspection of the integrals in addition to power counting
shows that the terms that contain a bosonic distribution function (in analogy
to the first terms in Eqs.\ (\ref{eq:3.18a}) and (\ref{eq:3.18b}),
respectively), have a potential to be of $O(\tau T \ln \delta)$ rather than of
$O(\tau T)$. However, the leading contribution to diagram (iii) does not
contain such terms. Diagrams (iv) and (v) do, but the logarithmic terms cancel
between these two diagrams, and this can be seen without performing the
integrals. We thus conclude
\bse
\label{eq:3.23}
\bea
\delta\sigma_{\perp} &=& \delta\sigma_{\perp}^{(iii)} + o(\tau T) = O(\tau T) +
o(\tau T),
\nonumber\\
\label{eq:3.23a}\\
\delta\sigma_{\text{L}} &=& \delta\sigma_{\text{L}}^{(iii)} +
\delta\sigma_{\text{L}}^{(iv,v)} + o(\tau T)
\nonumber\\
               &=& O(\tau T) + o(\tau T).
\label{eq:3.23b}
\eea
\ese

\paragraph{Diagram {\rm{(iii)}}}
\label{par:III.B.2.a}

After the above preliminary considerations, we now evaluate diagram (iii). The
leading contribution can be written
\bea
\delta\sigma_{ij}^{(iii)} &=& \frac{u_0 V_0}{8\pi}\,\frac{1}{T}
\int_{-\infty}^{\infty} \frac{d\epsilon}{\cosh^2(\epsilon/2T)} \frac{1}{V}
\sum_{\bm k} \int_{-\infty}^{\infty} \frac{du}{\pi}\,
\nonumber\\
&&\hskip -30pt \times n_{\text{F}}\left(\frac{u-\epsilon}{T}\right)\,
\chi''({\bm k},u)\, K_{ij}^{++-}\, L^{++,-}({\bm k}).
\label{eq:3.24}
\eea
Here
\bea
K_{ij}^{++-}(k) &=& \frac{1}{V} \sum_{\bm p} \frac{p_i\,p_j}{m_{\text{e}}^2}\
G_R({\bm p})\,G_R({\bm p})\,G_A({\bm p}) \nonumber\\
&=& -\delta_{ij}\,\frac{2\pi i}{3}\,\frac{k_{\text{F}}^2
N_{\text{F}}}{m_{\text{e}}^2}\, \tau^2 + O(\tau),
\label{eq:3.25}
\eea
and
\bea
L^{++,-}({\bm k}) &=& \frac{1}{V} \sum_{\bm p} \gamma({\bm k};{\bm p})\,
\gamma({\bm k};{\bm p}-{\bm k})\, G_R({\bm p})\, G_R({\bm p})\,
\nonumber\\
&&\hskip 90pt \times G_A({\bm p} - {\bm k})
\nonumber\\
&=& i\nu^2\, \frac{2\pi}{3}\, \frac{N_{\text{F}} m_{\text{e}}^2}{\lambda^2
k_{\text{F}}^2} + O(1/\tau,k_{\perp}^2).
\label{eq:3.26}
\eea
The second lines in Eqs.\ (\ref{eq:3.25}) and (\ref{eq:3.26}) are easy to
obtain in the AGD approximation. Only the term proportional to ${\bm
k}_{\perp}$ in $\gamma({\bm k};{\bm p})$, Eq.\ (\ref{eq:3.7e}), contributes to
the leading temperature dependence, hence the proportionality to $\nu^2$. We
again have dropped the frequency dependence of the Green functions, since it
does not contribute to the leading temperature dependence. Consequently, the
integral over $\epsilon$ in Eq.\ (\ref{eq:3.24}) can be performed. Using the
fact that the helimagnon spectrum $\chi''$ is an odd function of the frequency,
we can write
\bse
\label{eqs:3.27}
\bea
\delta\sigma_{ij}^{(iii)} &=& \frac{-u_0 V_0}{4\pi}\, \frac{1}{V} \sum_{\bm k}
\int_{-\infty}^{\infty} \frac{du}{\pi}\, \chi''({\bm k},u)\, C(u/2T)
\nonumber\\
&&\hskip 60pt \times K_{ij}^{++-}\, L^{++,-}({\bm k}),
\label{eq:3.27a}
\eea
with
\be
C(x) = \coth x - x/\sinh^2 x.
\label{eq:3.27b}
\ee
\ese

We next cast the expressions corresponding to diagrams (iv) and (v) in an
analogous form, before performing the final integrals.

\paragraph{Diagrams {\rm{(iv)}} and {\rm{(v)}}}
\label{par:III.B.2.a}

Using the same techniques as for diagram (iii), we find for the leading
contributions to diagrams (iv) and (v)
\bea
\delta\sigma_{ij}^{(iv)+ (v)} &=& \frac{-u_0 V_0}{2\pi}\, \frac{1}{V} \sum_{\bm
k} \int_{-\infty}^{\infty} \frac{du}{\pi}\, \chi''({\bm k},u)\, C(u/2T)
\nonumber\\
&&\hskip 30pt \times M_i^{+-,+}({\bm k})\, M_j^{+-,+}({\bm k}).
\label{eq:3.28}
\eea
Here
\bea
M_i^{+-,+}({\bm k}) &=& \frac{1}{V} \sum_{\bm p} \frac{p_i}{\me}\, \gamma({\bm
k};{\bm p})\, G_R({\bm p})\, G_A({\bm p})\, G_R({\bm p}-{\bm k})
\nonumber\\
&=& -\delta_{iz}\, \frac{2\pi}{3}\, \nu\, \frac{\NF}{\lambda}\, \tau +
O(\tau^0).
\label{eq:3.29}
\eea

\subsubsection{The conductivity in the ballistic limit}
\label{subsubsec:III.B.3}

Before we collect our results, we return to the question of the diagrams that
do not contribute to $O(\tau T)$ to lowest order in $q/k_{\text{F}}$. A
calculation shows that at the next order in $q/k_{\text{F}}$ they do
contribute, i.e., their behavior is the same as that of the diagrams we kept,
only the prefactor carries an additional factor of $(q/k_{\text{F}})^2$. For
diagram (vii) this is demonstrated in Appendix \ref{app:C}; for others, the
results are analogous. We note that if one wanted to keep these terms, one
would also have to take into account the difference between the quasiparticle
conductivity and the physical conductivity that was mentioned at the beginning
of Sec.\ \ref{sec:III}.

Collecting our results, we now have
\bea
\delta\sigma_{ij} &=& \frac{-u_0 V_0}{4\pi}\, \frac{1}{V} \sum_{\bm k}
\int_{-\infty}^{\infty} \frac{du}{\pi}\, \chi''({\bm k},u)\, C(u/2T)
\nonumber\\
&&\hskip 0pt \times \left[K_{ij}^{++,-}\,L^{++,-}({\bm k}) + 2 M_i^{+-,+}({\bm
k})\, M_j^{+-,+}({\bm k})\right].
\nonumber\\
\label{eq:3.30}
\eea
Here we show only terms that contribute to the leading temperature dependence
of $\delta\sigma$. With help of Eqs.\ (\ref{eq:3.25}, \ref{eq:3.26},
\ref{eq:3.29}, \ref{eq:3.27b}), and (\ref{eq:3.7d}) the final integrals are
easily performed. We find
\bse
\label{eqs:3.31}
\bea
\delta\sigma_{\perp} &=& \frac{\pi\nu^2}{576}\,\sigma_0\,
\left(\frac{q}{\kF}\right)^3\,
\left(\frac{\epsilon_{\text{F}}}{\lambda}\right)^2\,
\left[\frac{\Lambda}{2\epsilon_{\text{F}}} -
\frac{T}{\epsilon_{\text{F}}}\right],
\nonumber\\
\label{eq:3.31a}\\
\delta\sigma_{\text{L}} &=& 3\delta\sigma_{\perp}.
\label{eq:3.31b}
\eea
\ese
This result is valid in a temperature regime $T_{\text{ball}} < T < T_q$, as
explained below. $\Lambda$ is an ultraviolet energy cutoff, which must be
imposed, as in the Coulomb case,\cite{Zala_Narozhny_Aleiner_2001} since only
the hydrodynamic contributions to various parts of the integrands have been
kept. The cutoff-dependent part of $\delta\sigma$ is temperature independent;
it is an interaction correction to the Drude conductivity. The temperature
dependent part is independent of the cutoff. Notice that the constant
contribution to $\delta\sigma$ is positive, i.e., the effect of weak disorder
in conjunction with helimagnons is antilocalizing. Accordingly, the temperature
correction to the conductivity is negative.

A necessary condition for this result to be valid is that the parameter
$\delta$, Eq.\ (\ref{eq:3.21}), be small,
\be
T > T_{\text{ball}} \equiv \lambda/16 \sqrt{6}\,(\epsilon_{\text{F}}\tau)^2.
\label{eq:3.32}
\ee
The ballistic temperature scale $T_{\text{ball}}$ defined in this way marks the
lower temperature limit of the ballistic regime. Another necessary condition is
related to the fact the the helimagnon resonance frequency $\omega_0$ has the
form shown in Eq.\ (\ref{eq:1.5a}) only for wave numbers $k<q$. As was
explained in paper II, this defines another temperature scale $T_q = \lambda
q^2/6\kF^2$, and in order for Eqs.\ (\ref{eqs:3.31}) to hold we must have
$T<T_q$. This also identifies the order of magnitude of the UV cutoff: $\Lambda
= O(T_q)$. The temperature dependent contribution to $\delta\sigma$ in Eqs.\
(\ref{eqs:3.31}) is thus a small correction to the constant contribution. For
$k>q$ or, equivalently, $T>T_q$, the resonance frequency is $\omega_0({\bm k})
\propto \sqrt{c_{\perp}}\, k^2$, and hence the components of ${\bm k}$ scale as
$k_z \sim k_{\perp} \sim T^{1/2}$. Repeating the power counting arguments of
Sec.\ \ref{subsubsec:III.B.2} (see also Appendix \ref{app:B}), this yields, for
the temperature-dependent part of $\delta\sigma$,
\be
\delta\sigma_{\perp}(T) \propto \delta\sigma_{\text{L}}(T) \propto -\sigma_0\,
\left(\frac{q}{\kF}\right)^4\,
\left(\frac{\epsilon_{\text{F}}}{\lambda}\right)^{3/2}\,
\left(\frac{T}{\epsilon_{\text{F}}}\right)^{1/2},
\label{eq:3.33}
\ee
which is valid for $T>T_q$. We note that this is just Eqs.\ (\ref{eqs:3.31})
times $(T_q/T)^{1/2}$, so in the regime $T>T_q$, effectively a factor of
$\sqrt{T}$ gets replaced by $\sqrt{T_q}$.

We will discuss additional temperature scales, and the size of the ballistic
regime, in Sec.\ \ref{sec:IV} below.

\section{Discussion and Conclusion}
\label{sec:IV}

We now discuss our results. First, we give a detailed discussion of the range
of validity of our results, and of the various temperature scales involved. We
then give semi-quantitative estimates for the size of the predicted effects. In
evaluating these estimates, one should keep in mind that the qualitative
dependences on various parameters are accurate, and so are the ratios of
temperature scales etc., but the number-valued prefactors are model dependent
and should not be taken too seriously.

\subsection{Temperature scales, and relaxation rates}
\label{subsec:IV.A}

We start by giving an alternative derivation of the small parameter for the
weak-disorder expansion, Eq.\ (\ref{eq:3.21}). Consider the Green function $G$
given in Eq.\ (\ref{eq:3.12}). With ${\bm k}$ the soft wave number, and $u$ the
soft frequency, the ${\bm k}$-dependence of the various terms in the
perturbation theory is given by its spectrum,
\bse
\label{eqs:4.1}
\be
G''({\bm p}-{\bm k},u) = -\pi\,\Delta(u-\omega_1({\bm p}-{\bm k})),
\label{eq:4.1a}
\ee
with $\Delta$ a Lorentzian of the form
\be
\Delta(x) = \frac{1}{\pi}\,\frac{1/2\tau}{x^2 + 1/4\tau^2}\ .
\label{eq:4.1b}
\ee
\ese
In the clean limit, $\tau\rightarrow\infty$, $\Delta(x)$ turns into a delta
function. To determine the relevant scale, we recall the scaling of frequency
or temperature with the components of the wave vector. From Eq.\
(\ref{eq:1.5a}) we have
\be
u \sim T \sim \omega_0({\bm k}) \sim \sqrt{c_z}\,k_z \sim \sqrt{c_{\perp}}\,
k_{\perp}^2.
\label{eq:4.2}
\ee
With ${\bm p}$ on the Fermi surface, i.e., $\omega_1({\bm p})=0$, we have, from
Eq.\ (\ref{eq:3.6b}), $\omega_1({\bm p}-{\bm k}) = -{\bm p}_{\perp}\cdot{\bm
k}_{\perp}/\me + O(k_z)$. We now scale $u$ with $T$, $p$ with $\kF$, and
$k_{\perp}^2$ with $T/\sqrt{c_{\perp}}$. Keeping only leading terms, we can
write
\be
G''({\bm p}-{\bm k},u) = -\tau\delta\,\frac{\delta/2}{(\tilde{\bm p} \cdot
\tilde{\bm k}_{\perp})^2 + \delta^2/4}\ ,
\label{eq:4.3}
\ee
with $\tilde{\bm p}$ and $\tilde{\bm k}$ the scaled vectors ${\bm p}$ and ${\bm
k}$, respectively, and $\delta$ from Eq.\ (\ref{eq:3.21}). This confirms the
role of $\delta$ as the small parameter for the disorder expansion.

$\delta=1$ defines the temperature scale that was denoted by $T_{\text{ball}}$
in Sec.\ \ref{subsubsec:III.B.3}, and that we list here again:
\be
T_{\text{ball}} = \frac{1}{16 \sqrt{6}}\,
\frac{\lambda}{(\epsilon_{\text{F}}\tau)^2}\ .
\label{eq:4.4}
\ee
For a given disorder strength, this defines the lower temperature limit of the
ballistic regime.

A second relevant temperature scale is
\be
T_q = \lambda q^2/6\kF^2,
\label{eq:4.5}
\ee
which was introduced in paper II. As explained there, it is the energy scale
related to the crossover from the anisotropic helimagnon spectrum to an
isotropic ferromagnet-like spectrum with $\omega_0({\bm k}) \propto {\bm k}^2$.
$T_q > T_{\text{ball}}$ provided $\epsilon_{\text{F}}\tau > (3/8\sqrt{6})^{1/2}
\kF/q \approx 0.4\,\kF/q$. With $q/\kF \approx 0.02$, as is the case in MnSi,
this means $\epsilon_{\text{F}}\tau \agt 20$, which always holds for good
metals. The ballistic conductivity correction is then given by Eqs.\
(\ref{eqs:3.31}) in the temperature window $T_{\text{ball}} < T < T_q$, and by
Eq.\ (\ref{eq:3.33}) for $T>T_q$. For $T < T_{\text{ball}}$ one has diffusive
rather than ballistic transport behavior.

At this point it is useful to cast our result for the conductivity correction
in the form of a correction to the relaxation rate. The Drude conductivity plus
the ballistic correction is $\sigma = \sigma_0 + \delta\sigma_{\text{ball}}$,
which implies a total relaxation rate in the ballistic regime
\bea
\frac{1}{\tau} + \frac{1}{\tau_{\text{ball}}} &=& \frac{1}{\tau}\, \left(1 -
\frac{\delta\sigma_{\text{ball}}}{\sigma_0}\right)
\nonumber\\
&=& \frac{1}{\tau} + \frac{1}{\tau}\, \frac{\pi\nu^2}{1,152}\,
\left(\frac{q}{\kF}\right)^5\, \frac{\epsilon_{\text{F}}}{\lambda}\,
\frac{T}{T_q}\ .
\label{eq:4.6}
\eea
Here we have absorbed the constant contribution to the ballistic rate in the
Drude rate, and we have taken the correction to the longitudinal conductivity.
To obtain the total transport rate we also need to add the clean-limit rate
$1/\tau_1$ from paper II or Sec.\ \ref{subsec:III.A} above, which in the
current context is obtained from a ladder resummation of diagrams (i) and (ii)
in Fig.\ \ref{fig:5} and taking the clean limit. For $\nu=1$ we have
$g_1(\nu=1)\approx 0.2$, and from Eq.\ (\ref{eq:3.11c}) we find,
\be
\frac{1}{\tau_1} \approx \frac{\lambda}{50}\,\left(\frac{q}{\kF}\right)^8\,
\left(\frac{\epsilon_{\text{F}}}{\lambda}\right)^2\,
\left(\frac{T}{T_q}\right)^{5/2},
\label{eq:4.7}
\ee
for a total transport rate in the ballistic regime
\bea
\frac{1}{\tau_{\text{tr}}} &=& \frac{1}{\tau_1} + \frac{1}{\tau} +
\frac{1}{\tau_{\text{ball}}}
\nonumber\\
&\approx& \frac{\lambda}{50}\,\left(\frac{q}{\kF}\right)^8\,
\left(\frac{\epsilon_{\text{F}}}{\lambda}\right)^2\,
\left(\frac{T}{T_q}\right)^{5/2} + \frac{1}{\tau}
\nonumber\\
&& \hskip 50pt + \frac{1}{350\,\tau}\, \left(\frac{q}{\kF}\right)^5\,
\frac{\epsilon_{\text{F}}}{\lambda}\, \frac{T}{T_q}\ .
\label{eq:4.8}
\eea
Here we have put $\nu=1$ in Eq.\ (\ref{eq:4.6}) and have approximated the
numerical prefactor. It also is illustrative to recall the clean-limit
single-particle relaxation rate from paper II, which for generic wave vectors
is (see also Eq.\ (\ref{eq:3.9c})
\be
\frac{1}{\tau_{\text{clean}}^{\text{s.p.}}} \propto
\lambda\,\left(\frac{q}{\kF}\right)^6\,
\left(\frac{\epsilon_{\text{F}}}{\lambda}\right)^2\,
\left(\frac{T}{T_q}\right)^{3/2}.
\label{eq:4.9}
\ee
Notice that the clean transport rate is smaller than the clean single-particle
rate by a factor of $T/\lambda$, as was shown in paper II, whereas the
ballistic transport rate is qualitatively the same as the ballistic
single-particle rate, see Eqs.\ (\ref{eq:4.6}) and (III.3.12). That is, the
cancellation mechanism between self-energy contributions and vertex corrections
that is characteristic for clean transport problems (and also holds, e.g., in
the electron-phonon scattering problem) is not operative in the presence of
quenched disorder. As a result, $1/\tau_{\text{ball}}$ is small compared to
$1/\tau_{\text{clean}}^{\text{s.p.}}$ by a factor of $\delta$, Eq.\
(\ref{eq:3.21}), as was to be expected, but it is not necessarily small
compared to the clean transport rate $1/\tau_{\text{clean}}$. Rather, the
ballistic behavior will cross over to the clean behavior at a temperature
\be
T_{1-5/2} \approx 0.05 \lambda/(\epsilon_{\text{F}}\tau)^{2/3},
\label{eq:4.10}
\ee
which provides a third relevant temperature scale. A fourth one is given by the
temperature where the ballistic rate becomes equal to the clean Fermi-liquid
rate $1/\tau_{\text{FL}} = T^2/\epsilon_{\text{F}}$, which is
\be
T_{1-2} \approx \frac{10^{-3}}{\tau}\, \left(\frac{q}{\kF}\right)^3\,
\left(\frac{\epsilon_{\text{F}}}{\lambda}\right)^2.
\label{eq:4.11}
\ee

$T_q$, $T_{1-5/2}$, and $T_{1-2}$ all provide upper limits for the regime where
the conductivity correction is given by Eqs.\ (\ref{eqs:3.31}). We thus
conclude that the latter are valid in a temperature window given by
\be
T_{\text{ball}} < T < \text{Min}\left(T_q, T_{1-5/2}, T_{1-2}\right).
\label{eq:4.12}
\ee

Let us first compare $T_{1-5/2}$ with $T_{1-2}$ by writing the latter as
\be
T_{1-2} \approx T_{1-5/2}\, \frac{1}{50\,(\epsilon_{\text{F}}\tau)^{1/3}}\,
\left(\frac{q}{\kF}\right)^3\,
\left(\frac{\epsilon_{\text{F}}}{\lambda}\right)^3.
\label{eq:4.13}
\ee
If $\lambda = O(\epsilon_{\text{F}})$, then $T_{1-2} \ll T_{1-5/2}$. In a weak
helimagnet, where $\lambda/\epsilon_{\text{F}} \approx q/\kF$, this is still
true due to the small factor $1/50\,(\epsilon_{\text{F}}\tau)^{1/3} \ll 1$.

Similarly, we can write
\be
T_{1-2} \approx T_q\, \frac{1}{200\,\epsilon_{\text{F}}\tau}\,
\left(\frac{q}{\kF}\right)\,
\left(\frac{\epsilon_{\text{F}}}{\lambda}\right)^3.
\label{eq:4.14}
\ee
We again conclude that $T_{1-2}$ is the smaller of the two temperature scales
provided that
\be
\frac{\lambda}{\epsilon_{\text{F}}} \agt
\frac{1}{5\,(\epsilon_{\text{F}}\tau)^{1/3}}\,
\left(\frac{q}{\kF}\right)^{1/3}.
\label{eq:4.15}
\ee
As long as this condition is fulfilled, $T_{1-2}$ is the smallest of the three
lower bounds. We now compare $T_{1-2}$ and $T_{\text{ball}}$ by writing
\be
T_{1-2} \approx T_{\text{ball}}\, \frac{\epsilon_{\text{F}}\tau}{25}\,
\left(\frac{q}{\kF}\right)^3\,
\left(\frac{\epsilon_{\text{F}}}{\lambda}\right)^3.
\label{eq:4.16}
\ee
$T_{1-2}$ thus is larger than $T_{\text{ball}}$ provided
\be
\frac{\lambda}{\epsilon_{\text{F}}} \alt \frac{1}{3}\,
(\epsilon_{\text{F}}\tau)^{1/3}\, \frac{q}{\kF}\ .
\label{eq:4.17}
\ee
In order for the inequalities (\ref{eq:4.15}) and (\ref{eq:4.17}) to be
compatible, we must have
\be
\epsilon_{\text{F}}\tau > \kF/q,
\label{eq:4.18}
\ee
which is also roughly the condition for $T_q > T_{\text{ball}}$. This is not a
very stringent condition, and will generally be fulfilled in reasonably clean
systems.

We conclude that, if the inequalities (\ref{eq:4.18}) and (\ref{eq:4.17}) hold,
the ballistic conductivity correction is given by Eqs.\ (\ref{eqs:3.31}) in the
temperature regime $T_{\text{ball}} < T < T_{1-2}$. For lower temperatures the
behavior crosses over to diffusive transport, and for higher ones, to
Fermi-liquid behavior. If (\ref{eq:4.18}) holds, but (\ref{eq:4.17}) is
violated, then the Fermi-liquid $T^2$ behavior will mask the ballistic
$T$-dependence, and will have to be subtracted in order to observe the
ballistic effect.

We also need to remember that due to the broken rotational invariance in a
solid-state system, there actually is a term proportional to $k_{\perp}^2$
under the square root in Eq.\ (\ref{eq:1.5a}), but it has a small prefactor. As
was explained in papers I and II, this becomes relevant for temperatures below
a scale $T_{\text{so}} = T_q (q/\kF)^2$. In this context we further need to
come back to our discussion of the screening of the effective interaction given
in Eqs.\ (\ref{eqs:3.7}), see Ref.\ \onlinecite{screening_footnote}. As was
shown in Eqs.\ (III.2.25), screening modifies the temperature scale
$T_{\text{so}}$ to ${\tilde T}_{\text{so}} = T_q (q/\kF)^2 (q\vF/\lambda)^2$.
Requiring that this is temperature scale is smaller than $T_{\text{ball}}$
leads to one more constraint, namely,
\be
\epsilon_{\text{F}}\tau < (\kF/q)(\lambda/q\vF).
\label{eq:4.19}
\ee
Combined with (\ref{eq:4.18}) this leads to
\be
\kF/q < \epsilon_{\text{F}}\tau < (\kF/q)(\lambda/q\vF).
\label{eq:4.20}
\ee
as a necessary condition for the ballistic conductivity correction to be given
by Eqs.\ (\ref{eqs:3.31}). If the condition (\ref{eq:4.19}) is not fulfilled,
then the lower temperature limit of the behavior calculated above will be given
by ${\tilde T}_{\text{so}}$ rather than by $T{\text{ball}}$.

\subsection{Quantitative predictions for experiments}
\label{subsec:IV.B}

We now give some quantitative estimates, using parameter values relevant for
MnSi as follows (see paper I, and the references and discussion therein): $\kF
= 1.45\,\AA^{-1}$, $q/\kF = 0.024$, $\epsilon_{\text{F}} = 23,000\,{\text K}$,
$\me = 4m_0$, with $m_0$ the free-electron mass, $q\vF \approx 1,000\,{\text
K}$. The value of $\lambda$ is uncertain; the large magnetic moment suggests
that $\lambda$ is close to $\epsilon_{\text{F}}$, but it is possible that
$\lambda$ is smaller than $\epsilon_{\text{F}}$ by a factor of $40$. This
uncertainty in the value of the Stoner gap is a substantial impediment for
making experimental predictions, especially since the theory depends quite
strongly on whether $q\vF$ is larger or smaller than $\lambda$. Our
calculations are valid for $q\vF < \lambda$.

The residual resistivity of the cleanest samples in Ref.\
\onlinecite{Pfleiderer_Julian_Lonzarich_2001} was $\rho_0 \approx 0.33\,
\mu\Omega{\text{cm}}$, which corresponds to $\epsilon_{\text{F}}\tau \approx
1,000$. If $\lambda \approx \epsilon_{\text{F}}$, this is inside the disorder
window given by Eq.\ (\ref{eq:4.20}), and the condition (\ref{eq:4.18}) is
easily fulfilled. If $\lambda$ is substantially smaller than
$\epsilon_{\text{F}}$, then the second condition in Eq.\ (\ref{eq:4.20}) will
be violated, and the lower limit of the ballistic regime as calculated above
will be given by ${\tilde T}_{\text{so}}$ rather than by $T_{\text{ball}}$.
From Eqs.\ (\ref{eq:4.4}, \ref{eq:4.5}) we see that $T_{\text{ball}}$ is
smaller than $T_q$ by a factor of about 4,000, and from Eq.\ (\ref{eq:4.15}) we
see that $T_{1-2}$ is smaller than $T_q$ as long as
$\lambda/\epsilon_{\text{F}} \agt 0.005$, or $\lambda \agt 150\,\text{K}$.
Finally, from Eq.\ (\ref{eq:4.17}) it follows that $T_{1-2} > T_{\text{ball}}$
as long as $\lambda/\epsilon_{\text{F}} \alt 0.1$. We conclude that the
parameter values of MnSi provide a sizeable ballistic regime. However,
depending on the size of $\lambda$, it may be necessary to subtract the
Fermi-liquid $T^2$ contribution to the conductivity in order to observe the
ballistic correction.

The absolute size of the effect, on the other hand, is very small. From Eq.\
(\ref{eq:4.11}) we estimate that $T_{1-2}$ is at best, for the smallest
conceivable value of $\lambda$ ($\approx$ 500K), in the mK range, and from
Eqs.\ (\ref{eqs:3.31}) we have $\vert\delta\sigma_{\text{L}}/\sigma_0\vert
\approx 2\times 10^{-7}\, (\epsilon_{\text{F}}/\lambda)^2\,
T/\epsilon_{\text{F}}$. For $\lambda$ = 500K) this yields
$\vert\delta\sigma_{\text{L}}/\sigma_0\vert \approx 4\times 10^{-4}\,
T/\epsilon_{\text{F}}$. For temperatures on the order of $T_{1-2}$, this makes
for an extremely small effect, and even at $T\approx 1\,{\text{K}}$, which
requires subtraction of the Fermi-liquid contribution, the effect is small.

It is conceivable that in other materials the effect is larger, or that
artificial systems can be constructed, e.g., optical lattices, that have
parameter values leading to a larger effect. The most efficient way to increase
the effect would be a larger helix pitch wave number. In real systems, the
basic reason for the small absolute value of the effect is the prefactor
$(q/\kF)^3$ in Eqs.\ (\ref{eqs:3.31}). This in turn reflects the fact that any
effect of the helix will reflect the large size (on an atomic scale) of the
helix, which leads to correspondingly small energy scales. The same comment
holds for the conductivity in the clean case, see Eqs.\ (\ref{eq:3.11b},
\ref{eq:3.11c}). By contrast, if one writes the observed resistivity of MnSi in
the disordered phase as $\rho = \rho_0 [1 +
\text{const.}\times(T/\epsilon_{\text{F}})^{3/2}]$, then the experiment in
Ref.\ \onlinecite{Pfleiderer_Julian_Lonzarich_2001} yields $\text{const.} =
O(10^6)$. The anomalous temperature dependence of the resistivity in MnSi is
thus a very large effect that must be related to effects on small length
scales, and is not likely to be associated with remnants of helical order.

\subsection{Conclusion}
\label{subsec:IV.C}

In conclusion, we have applied the effective model for helimagnets that was
derived in paper III to determine the effects of helical magnetic order on the
electrical conductivity. In the clean limit, we reproduce the results obtained
earlier in paper II, but the effective model allows for a much simpler
calculation. We have applied this theory to determine the conductivity in the
ballistic regime, which in helimagnets is characterized by the requirement
$\sqrt{(\epsilon_{\text{F}}\tau)^2 T/\lambda} \gg 1$. Remarkably, we have found
that the temperature correction to the resistivity in bulk helimagnets is
linear in $T$, as it is in $2$-$d$ nonmagnetic metals. This analogy between
$3$-$d$ and $2$-$d$ systems is a consequence of the anisotropic dispersion
relation of the helical Goldstone mode or helimagnon. The absolute value of the
effect, with parameters values appropriate for known helimagnets, is very small
due to the large size of the helix. We finally mention that the transport
properties of helimagnets in the diffusive regime,
$\sqrt{(\epsilon_{\text{F}}\tau)^2 T/\lambda} \ll 1$, remain to be
investigated. Preliminary results suggest that they are less exotic than in the
ballistic regime, with no effective reduction of the
dimensionality.\cite{us_tbp}

\acknowledgments

This research was supported by the National Science Foundation under Grant Nos.
DMR-05-30314, DMR-05-29966, and PHY-05-51164. Part of this work was performed
at the Aspen Center for Physics.

\appendix

\section{Electrical resistivity due to a generic potential}
\label{app:A}

For pedagogical reasons, and to make a few technical points that are not
emphasized in the elementary literature, let us consider the resistivity of
nonmagnetic, spinless electrons due to scattering by an effective dynamical
potential. The familiar example of electron-phonon scattering, which leads to
the Bloch-Gr{\"u}neisen $T^5$ law, is a particular realization of this generic
case. The development in Sec.\ \ref{subsec:III.A} follows the same logic. The
only differences are that the resonance frequency of the bare Green function is
different, and that the potential depends on all three momenta in the
scattering process, not just on the net transferred momentum.

Consider spinless electrons interacting with a spin-independent,
frequency-dependent, effective potential $V({\bm k},i\Omega)$. We assume that
the spectrum of the potential, $V''({\bm k},u) = \text{Im}\,V({\bm k},i\Omega
\rightarrow u + i0)$, is soft at ${\bm k}=0$ and $u=0$. The Green function is
diagonal in both spin and momentum,
\be
{\cal G}({\bm p},i\omega) = \frac{1}{i\omega - \xi_{\bm p} - \Sigma({\bm
p},i\omega)}\ ,
\label{eq:A.1}
\ee\
where $\xi_{\bm p} = \epsilon_{\bm p} - \mu$, with $\epsilon_{\bm p}$ the
electronic energy-momentum relation, and $\mu$ the chemical potential. The
self-consistent Born equation for the self energy $\Sigma$, depicted
graphically in Fig.\ \ref{fig:3}, reads
\be
\Sigma({\bm p},i\omega) = -\frac{1}{V}\sum_{\bm p} T\sum_{i\Omega} V({\bm
k},i\Omega)\,{\cal G}({\bm p}-{\bm k},i\omega - i\Omega),
\label{eq:A.2}
\ee
and the integral equation for the vertex function ${\bm\Gamma}$ in a ladder
approximation, also shown in Fig.\ \ref{fig:3}, is
\bse
\label{eqs:A.3}
\begin{widetext}
\bea
{\bm\Gamma}({\bm p};i\omega,i\omega - i\Omega) &=& i{\bm j}({\bm p}) -
\frac{1}{V}\sum_{\bm k} T\sum_{i\Omega'} V({\bm k},i\Omega')\, {\cal G}({\bm
p}-{\bm k},i\omega - i\Omega')\, {\cal G}({\bm p}-{\bm k},i\omega - i\Omega' -
i\Omega)\,
\nonumber\\
&& \hskip 150pt \times {\bm\Gamma}({\bm p}-{\bm k};i\omega - i\Omega',i\omega -
i\Omega' - i\Omega).
\label{eq:A.3a}
\eea
If we define a scalar vertex function $\gamma$ by ${\bm\Gamma}({\bm p};i\omega,
i\omega - i\Omega) = i{\bm j}({\bm p})\, \gamma({\bm p};i\omega, i\omega -
i\Omega)$, we find that $\gamma$ obeys an integral equation
\bea
\gamma({\bm p};i\omega, i\omega - i\Omega) &=& 1 - \frac{1}{V}\sum_{\bm k}
T\sum_{i\Omega'} \frac{{\bm j}({\bm p})\cdot{\bm j}({\bm p}-{\bm k})}{{\bm
j}^2({\bm p})}\ V({\bm k},i\Omega')\, {\cal G}({\bm p}-{\bm k},i\omega -
i\Omega')\, {\cal G}({\bm p}-{\bm k},i\omega - i\Omega' - i\Omega)\,
\nonumber\\
&& \hskip 150pt \times {\bm\gamma}({\bm p}-{\bm k};i\omega - i\Omega',i\omega -
i\Omega' - i\Omega).
\label{eq:A.3b}
\eea
\end{widetext}
\ese
The polarization function and conductivity tensors are diagonal,
$\sigma_{ij}(i\Omega) = \delta_{ij}\, \sigma(i\Omega)$. The sum over fermionic
Matsubara frequencies in Eq.\ (\ref{eqs:A.3}) can be transformed into an
integral along the real axis by standard methods. This procedure yields two
terms where the frequency arguments of the Green functions lie on the same side
of the real axis, and two other terms where they lie on opposite sides. Only
the latter contribute to the leading result as the self energy goes to zero.
Since the real part of the self energy just renormalizes the Fermi energy, and
the imaginary part, which gives the relaxation rate, indeed goes to zero as
$T\to 0$, we need to keep only these retarded-advanced combinations for the
purpose of determining the leading low-temperature dependence of the
conductivity. The Kubo formula for the static conductivity $\sigma =
\lim_{\Omega\to 0} \Re \sigma(i\Omega \to \Omega + i0)$ becomes
\bea
\sigma &=& \frac{e^2}{3\pi} \int_{-\infty}^{\infty}
\frac{d\epsilon}{4T}\,\frac{1}{\cosh^2(\epsilon/2T)}\,\frac{1}{V}\sum_{\bm p}
\left({\bm j}({\bm p})\right)^2
\nonumber\\
&& \hskip 10pt \times \left\vert {\cal G}({\bm p},\epsilon + i0)\right\vert^2\,
\gamma({\bm p};\epsilon + i0,\epsilon - i0).
\label{eq:A.4}
\eea
The Green functions in Eq.\ (\ref{eq:A.4}) ensure that the dominant
contribution to the sum over wave vectors in the limit of a vanishing self
energy comes from ${\bm p}$ such that $\xi_{\bm p} = \epsilon$. Furthermore,
since $\epsilon$ scales with $T$, for the leading temperature dependence we can
neglect all $\epsilon$-dependencies that do not occur in the form $\epsilon/T$.
In a nearly-free electron model, with a spherical Fermi surface with Fermi wave
number $k_{\text{F}}$, and ${\bm j}({\bm p}) = {\bm p}/m_{\text{e}}$ with
$m_{\text{e}}$ the effective electron mass, we thus have
\bse
\label{eqs:A.5}
\be
\sigma = \frac{e^2 \kF^2}{3\me^2} \int_{-\infty}^{\infty}
\frac{d\epsilon}{4T}\,\frac{1}{\cosh^2(\epsilon/2T)}\,
\frac{\Lambda({\epsilon})}{\Gamma_0({\epsilon})}\ .
\label{eq:A.5a}
\ee
Here we have defined
\bea
\Lambda(\epsilon) &\equiv& \frac{1}{V}\sum_{\bm p} \delta(\xi_{\bm p})\,
\gamma({\bm p};\epsilon + i0,\epsilon - i0),
\label{eq:A.5b}\\
\Gamma_0(\epsilon) &\equiv& \frac{-1}{V}\sum_{\bm p} \delta(\xi_{\bm p})\,
\Im\Sigma({\bm p},\epsilon + i0),
\label{eq:A.5c}
\eea
\ese
and we neglect the real part of the self energy, which only redefines the zero
of energy.

Using analogous arguments, we find from Eq.\ (\ref{eq:A.3b}) that $\Lambda$
obeys an integral equation
\bea
\Lambda(\epsilon) &=& 1 - N_{\text{F}}\int du\,{\bar V}''(u)\,\left[
n_{\text{B}}\left(\frac{u}{T}\right) + n_{\text{F}}\left(\frac{\epsilon +
u}{T}\right)\right]\,
\nonumber\\
&&\hskip 60pt \times \Lambda(\epsilon + u)/\Gamma_0(\epsilon + u).
\label{eq:A.6}
\eea
Here $n_{\text{B}}(x) = 1/(e^x -1)$ and $n_{\text{F}}(x) = 1/(e^x + 1)$ are the
Bose and Fermi distribution function, respectively, and
\bse
\label{eqs:A.7}
\be
{\bar V}''(u) = \frac{1}{S_{\text{F}}^2}\,\frac{1}{V^2}\sum_{{\bm k},{\bm p}}
\delta(\xi_{\bm k})\,\delta(\xi_{\bm p})\,V''({\bm k}-{\bm p},u)\,{\bm
k}\cdot{\bm p}/k^2
\label{eq:A.7a}
\ee
with $S_{\text{F}} = (1/V)\sum_{\bm k} \delta(\xi_{\bm k})$, is an $l=1$
average of the spectrum of the potential over the Fermi surface. For the
purpose of finding the leading temperature dependence of the conductivity, it
can be written
\be
{\bar V}''(u) = {\bar V}_0''(u) - {\bar V}_1''(u),
\label{eq:A.7b}
\ee
with
\be
{\bar V}_n''(u) = \frac{1}{2k_{\text{F}}^2}
\int_0^{2k_{\text{F}}}dp\,p\,\left(p^2/2k_{\text{F}}^2\right)^n\,V''(p,u).
\label{eq:A.7c}
\ee
\ese

The integral equation (\ref{eq:A.6}) is not easy to solve. However, in an
approximation that replaces $\Lambda(\epsilon + u)/\Gamma_0(\epsilon + u)$
under the integral by $\Lambda(\epsilon)/\Gamma_0(\epsilon)$, it turns into an
algebraic equation whose solution is
\bse
\label{eqs:A.8}
\be
\Lambda(\epsilon) = \Gamma_0(\epsilon)/\Gamma_1(\epsilon),
\label{eq:A.8a}
\ee
Here we have used the fact that, in the limit of a small self energy, one find
from Eqs.\ (\ref{eq:A.5c}) and (\ref{eq:A.2}),
\be
\Gamma_0(\epsilon) = -N_{\text{F}} \int du\,{\bar
V}_0''(u)\,\left[n_{\text{B}}\left(\frac{u}{T}\right) +
n_{\text{F}}\left(\frac{\epsilon + u}{T}\right)\right],
\label{eq:A.8b}
\ee
and we have defined
\be
\Gamma_1(\epsilon) = -N_{\text{F}} \int du\,{\bar
V}_1''(u)\,\left[n_{\text{B}}\left(\frac{u}{T}\right) +
n_{\text{F}}\left(\frac{\epsilon + u}{T}\right)\right].
\label{eq:A.8c}
\ee
\ese

We see that the vertex function $\Lambda$ effective replaces the
single-particle relaxation rate $\Gamma$ with the transport relaxation rate
$\Gamma_1$. To see the relation between the two, we recall that the frequency
$u$ scales with the temperature. For potentials where the frequency scales with
some (positive) power of the wave number, $\Gamma_1$ will thus depend on a
higher power of the temperature as $T\to 0$ than $\Gamma$. As an example,
consider the case of electron scattering by acoustic phonons, where $V''(p,u)
\propto cp\,[\delta(u-cp) - \delta(u+cp)]$, with $c$ the speed of sound. In
this case, $\Gamma_0(\epsilon) \propto T^3\,\gamma_0(\epsilon/T)$, whereas
$\Gamma_1(\epsilon) \propto T^5\,\gamma_1(\epsilon/T)$, where
\be
\gamma_n(y) = \int_0^{\infty} dx\,x^{2(n+1)}\left[2n_{\text{B}}(x) +
n_{\text{F}}(x+y) + n_{\text{F}}(x-y)\right].
\label{eq:A.9}
\ee
In this case, the single-particle scattering rate shows a $T^3$ dependence,
whereas the transport scattering rate, and hence the resistivity, display the
familiar Bloch-Gr{\"u}neisen law, $\sigma \propto T^5$.

Most of the technical development sketched above can be found in
textbooks.\cite{Mahan_1981} What is usually not stressed is the fact that the
approximate solution, Eq.\ (\ref{eq:A.8a}), of the integral equation,
(\ref{eq:A.6}), yields the asymptotically exact temperature dependence
(although not the prefactor) of the conductivity. The fact that it does has, to
our knowledge, never been established within diagrammatic many-body theory (and
it is not proven by the above arguments), but it can been seen from the fact
that the asymptotic solution reproduces the lowest-order variational solution
of the Boltzmann equation.\cite{Wilson_1954} The relation between a
diagrammatic evaluation of the Kubo formula and solutions of the Boltzmann
equation is complex, and will be discussed in more detail
elsewhere.\cite{us_tbp}

\section{Power counting for diagrams $\text{(i)}$ and $\text{(ii)}$}
\label{app:B}

Here we provide the arguments that lead to Eq.\ (\ref{eq:3.20}). We first do a
power-counting analysis of Eqs.\ (\ref{eqs:3.18}). From Eqs.\ (\ref{eq:1.5a},
\ref{eq:3.7c}, \ref{eqs:3.18}) we see that the soft helimagnon wave number $k$
scales with temperature as $k_z \sim k_{\perp}^2 \sim T$. The frequencies scale
as $u\sim\epsilon\sim T$, and $\chi''({\bm k},u) \sim 1/T^2$. Consequently, the
conductivity corrections $\delta\sigma^{(i,ii)}$ scale as $\delta\sigma \sim
TJ$ for a given integrand $J(k)$ (or $I(k)$).

First consider the integral $J_{ij}^{++-,+}(k)$, Eq.\ (\ref{eq:3.19a}). For
power-counting purposes, the integration variable ${\bm p}$ scales as $T^0$,
and the leading term in the vertex $\gamma$ scales as $\gamma({\bm k},{\bm p})
\sim k_{\perp} \sim T^{1/2}$. A representation that suffices for power counting
is thus
\bse
\label{eqs:B.1}
\begin{widetext}
\be
J_{ij}^{++-,+}(k) \propto k_{\perp}^2 \int_{-\infty}^{\infty} d\xi
\int_{-1}^{1} d\eta\, \frac{1}{(\xi - i/2\tau)^2}\,\frac{1}{\xi + i/2\tau}\,
\frac{1}{\xi - i/2\tau - v_{\text{F}}k \eta} \propto
\frac{\tau^2\,k_{\perp}^2}{k} \int_0^{v_{\text F}k\tau} \frac{dx}{1+x^2}
\label{eq:B.1a}
\ee
\end{widetext}
in the AGD approximation. For $v_{\text{F}}k \gg 1/\tau$ we thus have
$J_{ij}^{++-,+}(k) \propto \tau^2 k_{\perp}^2/k$, with corrections carrying an
extra factor of $1/v_{\text{F}}k\tau \sim 1/\tau T^{1/2}$, or
\be
J_{ij}^{++-,+}(k) \sim \tau^2\,T^{1/2} + \tau,
\label{eq:B.1b}
\ee
\ese
which leads to $\delta\sigma \propto \tau^2 T^{3/2} + \tau T$. Analogous
arguments yield
\bea
J_{ij}^{++-,-}(k) &\sim& \tau^2\,T^{1/2} + \tau,
\label{eq:B.2}\\
J_{ij}^{+-,+-}(k) &\sim& \tau^2\,T^{1/2} + \tau,
\label{eq:B.3}\\
J_{ij}^{+-,++}(k) &\sim& \tau.
\label{eq:B.4}
\eea
The convolutions $I$, compared to the corresponding $J$, carry an additional
factor of $k_{\perp} \sim T^{1/2}$. In addition, the resulting vector nature of
the integrand leads to an another factor of either $k_{\perp} \sim T^{1/2}$, or
$k_z \sim T$. Therefore, the $I$ carry an additional factor of $T$ compared to
the corresponding $J$. Terms that were dropped in writing Eqs.\
(\ref{eqs:3.18}) involved $J^{+++,+}$, $J^{+++,-}$, $J^{++,++}$, and
$J^{++,--}$, which are of higher order in the disorder by at least three powers
of $1/\tau$. Including terms of $O(\tau T)$, we thus can write the conductivity
correction, Eqs.\ (\ref{eqs:3.18}),
\begin{widetext}
\bea
\delta\sigma^{(i)}_{ij} + \delta\sigma^{(ii)}_{ij} &=&
\frac{-V_0}{4\pi}\,\frac{1}{T} \int_{-\infty}^{\infty}
\frac{d\epsilon}{\cosh^2(\epsilon/T)}\,\frac{1}{V} \sum_{\bm k}
\int_{-\infty}^{\infty} \frac{du}{\pi}\, \chi''({\bm k},u)\
\Biggl[n_{\text{B}}\left(\frac{u}{T}\right)\, \text{Re}\,\left[
J_{ij}^{++-,+}(k) + \frac{1}{2}\,J_{ij}^{+-,+-}(k)\right]
\nonumber\\
&& \hskip 50pt +
\frac{1}{2}\,n_{\text{F}}\left(\frac{u-\epsilon}{T}\right)\,\text{Re}\, \left[
J_{ij}^{++-,+}(k) + J_{ij}^{+-,+-}(k) - J_{ij}^{++-,-}(k) - J_{ij}^{+-,++}(k)
\right] \Biggr].
\label{eq:B.5}
\eea
The $J$ can be simplified by means of partial fraction decompositions. For the
relevant combinations one finds
\bse
\label{eqs:B.6}
\bea
\text{Re}\,\left[ J_{ij}^{++-,+}(k) + \frac{1}{2}\,J_{ij}^{+-,+-}(k)\right]
\propto \tau^2\,k_{\perp}^3 \sim \tau^2\,T^{3/2},
\label{eq:B.6a}\\
\text{Re}\, \left[ J_{ij}^{++-,+}(k) + J_{ij}^{+-,+-}(k) - J_{ij}^{++-,-}(k) -
J_{ij}^{+-,++}(k) \right] = o(\tau\,T^0).
\label{eq:B.6b}
\eea
\ese
\end{widetext}
This leads to Eq.\ (\ref{eq:3.20}).

\section{Diagram $\text{(vii)}$}
\label{app:C}

Here we consider diagram (vii) in Fig.\ \ref{fig:5}(c) as a prototype of a
class of diagrams that do not contribute to the leading behavior of the
conductivity if evaluated to lowest order in $q/k_{\text{F}}$. The leading
contribution to the conductivity correction from this diagram can be written
\begin{widetext}
\bea
\delta\sigma_{ij}^{(vii)} &=& \frac{-u_0\,V_0}{8\pi\me^2}\, \frac{1}{T}
\int_{-\infty}^{\infty} \frac{d\epsilon}{\cosh^2(\epsilon/2T)}\,
\int_{-\infty}^{\infty} \frac{du}{\pi}\,
n_{\text{F}}\left(\frac{u-\epsilon}{T}\right)\, \text{Im}\, \frac{1}{V}
\sum_{\bm k} \chi_R({\bm k},u)\, \frac{1}{V} \sum_{\bm p} \gamma({\bm k},{\bm
p})\, G_R({\bm p})\,G_A({\bm p}-{\bm k})\, \nonumber\\
&&\hskip 150pt \times \frac{1}{V} \sum_{{\bm p}'} p'_i p'_j\,\gamma({\bm
k},{\bm p}')\, G_R({\bm p}')\, G_R({\bm p}')\, G_A({\bm p}')\, G_A({\bm
p}'-{\bm k}),
\label{eq:C.1}
\eea
\end{widetext}
which shows the (2,4) structure mentioned in Sec.\ \ref{subsubsec:III.B.2}. The
bosonic distribution function does not contribute to this diagram, so it can be
at most of $O(\tau T)$. With the convolutions evaluated for $q=0$, power
counting shows that it is of $O(\tau T^2)$, and an explicit calculation
confirms this. Now we expand the resonance frequency $\omega_1({\bm p})$, Eq.\
(\ref{eq:3.6b}), to first order in $q$: $\omega_1({\bm p}) = \xi_{\bm p} + {\bm
p}\cdot{\bm q}/2m_{\text{e}} + O(q^2)$. For the leading contribution to the
first convolution in Eq.\ (\ref{eq:C.1}) we then find
\bea
N^{+,-}({\bm k}) &\equiv& \frac{1}{V} \sum_{\bm p} \gamma({\bm k},{\bm p})\,
G_R({\bm p})\,G_A({\bm p}-{\bm k}) \nonumber\\
&\propto& \frac{\nu}{\lambda k_{\text{F}}^2}\, \left[\frac{k_{\perp}^2
k_z}{k^3} + \frac{k_{\perp}^2 q}{k^2 k_{\text{F}}} + O(q^2)\right].
\label{eq:C.2}
\eea
We see that, at linear order in $q$, a factor that used to be $k_z/k \sim
k_z/k_{\perp} \sim T^{1/2}$ gets replaced by $q/k_{\text{F}} \sim T^0$. The
same holds for the other convolution. As a result, the diagram is of $O(\tau
T)$, and an explicit calculation shows that the dependence of the prefactor on
$\epsilon_{\text{F}}/\lambda$ and $q/k_{\text{F}}$ are the same as for diagram
(iii), with the exception of the additional factor of $(q/k_{\text{F}})^2$. We
thus have
\be
\delta\sigma_{ij}^{(vii)} \propto \sigma_0\,\nu^2\,
\left(\frac{\epsilon_{\text{F}}}{\lambda}\right)^2\,
\left(\frac{q}{k_{\text{F}}}\right)^5\, \frac{T}{\epsilon_{\text{F}}}\ .
\label{eq:C.3}
\ee


\begin{thebibliography}{25}
\expandafter\ifx\csname natexlab\endcsname\relax\def\natexlab#1{#1}\fi
\expandafter\ifx\csname bibnamefont\endcsname\relax
  \def\bibnamefont#1{#1}\fi
\expandafter\ifx\csname bibfnamefont\endcsname\relax
  \def\bibfnamefont#1{#1}\fi
\expandafter\ifx\csname citenamefont\endcsname\relax
  \def\citenamefont#1{#1}\fi
\expandafter\ifx\csname url\endcsname\relax
  \def\url#1{\texttt{#1}}\fi
\expandafter\ifx\csname urlprefix\endcsname\relax\def\urlprefix{URL }\fi
\providecommand{\bibinfo}[2]{#2} \providecommand{\eprint}[2][]{\url{#2}}

\bibitem[{\citenamefont{Gorkov et~al.}(1979)\citenamefont{Gorkov, Larkin, and
  Khmelnitskii}}]{Gorkov_Larkin_Khmelnitskii_1979}
\bibinfo{author}{\bibfnamefont{L.~P.} \bibnamefont{Gorkov}},
  \bibinfo{author}{\bibfnamefont{A.}~\bibnamefont{Larkin}}, \bibnamefont{and}
  \bibinfo{author}{\bibfnamefont{D.~E.} \bibnamefont{Khmelnitskii}},
  \bibinfo{journal}{Pis'ma Zh. Eksp. Teor. Fiz.} \textbf{\bibinfo{volume}{30}},
  \bibinfo{pages}{248} (\bibinfo{year}{1979}), \bibinfo{note}{[JETP Lett. {\bf
  30}, 228 (1979)]}.

\bibitem[{not()}]{notation_footnote}
\bibinfo{note}{Troughout this paper we use a notation where $x\propto y$ stands
  for ``$x$ is proportional to $y$'', and $x\sim y$ stands for ``$x$ scales as
  $y$''}.

\bibitem[{\citenamefont{Abrahams et~al.}(1979)\citenamefont{Abrahams, Anderson,
  Licardello, and Ramakrishnan}}]{Abrahams_et_al_1979}
\bibinfo{author}{\bibfnamefont{E.}~\bibnamefont{Abrahams}},
  \bibinfo{author}{\bibfnamefont{P.~W.} \bibnamefont{Anderson}},
  \bibinfo{author}{\bibfnamefont{D.~C.} \bibnamefont{Licardello}},
  \bibnamefont{and} \bibinfo{author}{\bibfnamefont{T.~V.}
  \bibnamefont{Ramakrishnan}}, \bibinfo{journal}{Phys. Rev. Lett.}
  \textbf{\bibinfo{volume}{42}}, \bibinfo{pages}{673} (\bibinfo{year}{1979}).

\bibitem[{\citenamefont{Lee and Ramakrishnan}(1985)}]{Lee_Ramakrishnan_1985}
\bibinfo{author}{\bibfnamefont{P.~A.} \bibnamefont{Lee}} \bibnamefont{and}
  \bibinfo{author}{\bibfnamefont{T.~V.} \bibnamefont{Ramakrishnan}},
  \bibinfo{journal}{Rev. Mod. Phys.} \textbf{\bibinfo{volume}{57}},
  \bibinfo{pages}{287} (\bibinfo{year}{1985}).

\bibitem[{\citenamefont{Bergmann}(1984)}]{Bergmann_1984}
\bibinfo{author}{\bibfnamefont{G.}~\bibnamefont{Bergmann}},
  \bibinfo{journal}{Phys. Rep.} \textbf{\bibinfo{volume}{101}},
  \bibinfo{pages}{1} (\bibinfo{year}{1984}).

\bibitem[{\citenamefont{Schmid}(1974)}]{Schmid_1974}
\bibinfo{author}{\bibfnamefont{A.}~\bibnamefont{Schmid}}, \bibinfo{journal}{Z.
  Phys. B} \textbf{\bibinfo{volume}{271}}, \bibinfo{pages}{251}
  (\bibinfo{year}{1974}). \bibinfo{note}{This paper gives the result at $T=0$
  as a function of the distance $E$ from the Fermi surface, which is $1/\tau_C
  = (\pi/8)\,E^2/\epsilon_{\text{F}}$.}

\bibitem[{\citenamefont{Altshuler et~al.}(1980)\citenamefont{Altshuler, Aronov,
  and Lee}}]{Altshuler_Aronov_Lee_1980}
\bibinfo{author}{\bibfnamefont{B.~L.} \bibnamefont{Altshuler}},
  \bibinfo{author}{\bibfnamefont{A.~G.} \bibnamefont{Aronov}},
  \bibnamefont{and} \bibinfo{author}{\bibfnamefont{P.~A.} \bibnamefont{Lee}},
  \bibinfo{journal}{Phys. Rev. Lett.} \textbf{\bibinfo{volume}{44}},
  \bibinfo{pages}{1288} (\bibinfo{year}{1980}).

\bibitem[{\citenamefont{Altshuler and Aronov}(1984)}]{Altshuler_Aronov_1984}
\bibinfo{author}{\bibfnamefont{B.~L.} \bibnamefont{Altshuler}}
  \bibnamefont{and} \bibinfo{author}{\bibfnamefont{A.~G.}
  \bibnamefont{Aronov}}, \emph{\bibinfo{title}{Electron-Electron Interactions
  in Disordered Systems}} (\bibinfo{publisher}{North-Holland, Amsterdam},
  \bibinfo{year}{1984}), \bibinfo{note}{edited by M. Pollak and A.~L. Efros}.

\bibitem[{\citenamefont{Zala et~al.}(2001)\citenamefont{Zala, Narozhny, and
  Aleiner}}]{Zala_Narozhny_Aleiner_2001}
\bibinfo{author}{\bibfnamefont{G.}~\bibnamefont{Zala}},
  \bibinfo{author}{\bibfnamefont{B.~N.} \bibnamefont{Narozhny}},
  \bibnamefont{and} \bibinfo{author}{\bibfnamefont{I.~L.}
  \bibnamefont{Aleiner}}, \bibinfo{journal}{Phys. Rev. B}
  \textbf{\bibinfo{volume}{64}}, \bibinfo{pages}{214204}
  (\bibinfo{year}{2001}).

\bibitem[{\citenamefont{Belitz et~al.}(2006{\natexlab{a}})\citenamefont{Belitz,
  Kirkpatrick, and Rosch}}]{Belitz_Kirkpatrick_Rosch_2006a}
\bibinfo{author}{\bibfnamefont{D.}~\bibnamefont{Belitz}},
  \bibinfo{author}{\bibfnamefont{T.~R.} \bibnamefont{Kirkpatrick}},
  \bibnamefont{and} \bibinfo{author}{\bibfnamefont{A.}~\bibnamefont{Rosch}},
  \bibinfo{journal}{Phys. Rev. B} \textbf{\bibinfo{volume}{73}},
  \bibinfo{pages}{054431} (\bibinfo{year}{2006}{\natexlab{a}}),
  \bibinfo{note}{(paper I)}.

\bibitem[{\citenamefont{Belitz et~al.}(2006{\natexlab{b}})\citenamefont{Belitz,
  Kirkpatrick, and Rosch}}]{Belitz_Kirkpatrick_Rosch_2006b}
\bibinfo{author}{\bibfnamefont{D.}~\bibnamefont{Belitz}},
  \bibinfo{author}{\bibfnamefont{T.~R.} \bibnamefont{Kirkpatrick}},
  \bibnamefont{and} \bibinfo{author}{\bibfnamefont{A.}~\bibnamefont{Rosch}},
  \bibinfo{journal}{Phys. Rev. B} \textbf{\bibinfo{volume}{74}},
  \bibinfo{pages}{024409} (\bibinfo{year}{2006}{\natexlab{b}}),
  \bibinfo{note}{(paper II). See also the Erratum published as Phys. Rev. B
  {\bf 76}, 149902 (2007). A version that incorporates the Erratum, and
  corrects typos in the published paper, is available as
  arXiv:cond-mat/0604427}.

\bibitem[{F_f()}]{F_footnote}
\bibinfo{note}{Due to the Stoner splitting, one strictly speaking has to
  distinguish between Fermi-surface properties, such as the Fermi wave number,
  the density of states at the Fermi surface, etc., in the two Stoner bands.
  For a weak helimagnet the differences between these quantities are small, and
  we will systematically neglect them. This amounts to working to lowest order
  in the small parameter $\lambda/\epsilon_{\text{F}}$.}

\bibitem[{\citenamefont{Ishikawa et~al.}(1976)\citenamefont{Ishikawa, Tajima,
  Bloch, and Roth}}]{Ishikawa_et_al_1976}
\bibinfo{author}{\bibfnamefont{Y.}~\bibnamefont{Ishikawa}},
  \bibinfo{author}{\bibfnamefont{K.}~\bibnamefont{Tajima}},
  \bibinfo{author}{\bibfnamefont{D.}~\bibnamefont{Bloch}}, \bibnamefont{and}
  \bibinfo{author}{\bibfnamefont{M.}~\bibnamefont{Roth}},
  \bibinfo{journal}{Solid State Commun.} \textbf{\bibinfo{volume}{19}},
  \bibinfo{pages}{525} (\bibinfo{year}{1976}).

\bibitem[{\citenamefont{Pfleiderer et~al.}(1997)\citenamefont{Pfleiderer,
  McMullan, Julian, and Lonzarich}}]{Pfleiderer_et_al_1997}
\bibinfo{author}{\bibfnamefont{C.}~\bibnamefont{Pfleiderer}},
  \bibinfo{author}{\bibfnamefont{G.~J.} \bibnamefont{McMullan}},
  \bibinfo{author}{\bibfnamefont{S.~R.} \bibnamefont{Julian}},
  \bibnamefont{and} \bibinfo{author}{\bibfnamefont{G.~G.}
  \bibnamefont{Lonzarich}}, \bibinfo{journal}{Phys. Rev. B}
  \textbf{\bibinfo{volume}{55}}, \bibinfo{pages}{8330} (\bibinfo{year}{1997}).

\bibitem[{\citenamefont{Pfleiderer et~al.}(2001)\citenamefont{Pfleiderer,
  Julian, and Lonzarich}}]{Pfleiderer_Julian_Lonzarich_2001}
\bibinfo{author}{\bibfnamefont{C.}~\bibnamefont{Pfleiderer}},
  \bibinfo{author}{\bibfnamefont{S.~R.} \bibnamefont{Julian}},
  \bibnamefont{and} \bibinfo{author}{\bibfnamefont{G.~G.}
  \bibnamefont{Lonzarich}}, \bibinfo{journal}{Nature}
  \textbf{\bibinfo{volume}{414}}, \bibinfo{pages}{427} (\bibinfo{year}{2001}).

\bibitem[{pap({\natexlab{a}})}]{paper_III}
\bibinfo{note}{T.~R. Kirkpatrick, D. Belitz, and Ronojoy Saha, arXiv:0806.0614
  (paper III).}

\bibitem[{pap({\natexlab{b}})}]{paper_II_footnote}
\bibinfo{note}{References to paper II refer to the electronic version available
  as arXiv:0604.427, which incorporates the Erratum, and corrects typographic
  errors and minor mistakes in the published version.}

\bibitem[{\citenamefont{Baym and Kadanoff}(1961)}]{Baym_Kadanoff_1961}
\bibinfo{author}{\bibfnamefont{G.}~\bibnamefont{Baym}} \bibnamefont{and}
  \bibinfo{author}{\bibfnamefont{L.~P.} \bibnamefont{Kadanoff}},
  \bibinfo{journal}{Phys. Rev.} \textbf{\bibinfo{volume}{124}},
  \bibinfo{pages}{287} (\bibinfo{year}{1961}).

\bibitem[{\citenamefont{Kadanoff and Baym}(1962)}]{Kadanoff_Baym_1962}
\bibinfo{author}{\bibfnamefont{L.~P.} \bibnamefont{Kadanoff}} \bibnamefont{and}
  \bibinfo{author}{\bibfnamefont{G.}~\bibnamefont{Baym}},
  \emph{\bibinfo{title}{Quantum Statistical Mechanics}}
  (\bibinfo{publisher}{W.A. Benjamin, New York}, \bibinfo{year}{1962}).

\bibitem[{scr()}]{screening_footnote}
\bibinfo{note}{This is the unscreened effective potential. As explained in Sec.
  II.E of paper III, screening of the helimagnon susceptibility $\chi$
  effective produces a term of $O({\bm k}_{\perp}^2)$ in the helimagnon
  frequency $\omega_0$, which has a small prefactor and is comparable to a term
  that is generated by crystal-field effects that break the rotational
  invariance of our model. For simplicity, we ignore this effect for the time
  being, and will return to it and estimate its magnitude in Sec.\
  \ref{subsec:IV.A} below.}

\bibitem[{per()}]{perturbation_footnote}
\bibinfo{note}{A proof of this statement would require renormalization-group
  arguments, which in turn require a field-theoretic formulation of the problem
  that does not currently exist. However, the structure of the perturbation
  theory developed here strongly suggests that it is true, and it is believed
  to be true for the case of electrons interacting via a Coulomb interaction.}

\bibitem[{\citenamefont{Abrikosov et~al.}(1963)\citenamefont{Abrikosov, Gorkov,
  and Dzyaloshinski}}]{Abrikosov_Gorkov_Dzyaloshinski_1963}
\bibinfo{author}{\bibfnamefont{A.~A.} \bibnamefont{Abrikosov}},
  \bibinfo{author}{\bibfnamefont{L.~P.} \bibnamefont{Gorkov}},
  \bibnamefont{and} \bibinfo{author}{\bibfnamefont{I.~E.}
  \bibnamefont{Dzyaloshinski}}, \emph{\bibinfo{title}{Methods of Quantum Field
  Theory in Statistical Physics}} (\bibinfo{publisher}{Dover, New York},
  \bibinfo{year}{1963}).

\bibitem[{us_()}]{us_tbp}
\bibinfo{note}{T.~R. Kirkpatrick and D. Belitz, unpublished.}

\bibitem[{\citenamefont{Mahan}(1981)}]{Mahan_1981}
\bibinfo{author}{\bibfnamefont{G.~D.} \bibnamefont{Mahan}},
  \emph{\bibinfo{title}{Many-Particle Physics}} (\bibinfo{publisher}{Plenum,
  New York}, \bibinfo{year}{1981}).

\bibitem[{\citenamefont{Wilson}(1954)}]{Wilson_1954}
\bibinfo{author}{\bibfnamefont{A.~H.} \bibnamefont{Wilson}},
  \emph{\bibinfo{title}{The Theory of Metals}} (\bibinfo{publisher}{Cambridge
  University Press, Cambridge}, \bibinfo{year}{1954}).

\end{thebibliography}

\end{document}